 \documentclass[twocolumn]{aastex631}

\begin{document}

\title{Measuring Hubble Constant with Dark Neutron Star-Black Hole Mergers}

\author[0000-0003-1604-9805]{Banafsheh Shiralilou}
\affiliation{GRAPPA, Anton Pannekoek Institute for Astronomy and Institute of High-Energy
Physics, University of Amsterdam, Science Park 904, 1098 XH Amsterdam, The
Netherlands}
\author[0000-0002-9397-786X]{Geert Raaiijmakers}
\affiliation{GRAPPA, Anton Pannekoek Institute for Astronomy and Institute of High-Energy
Physics, University of Amsterdam, Science Park 904, 1098 XH Amsterdam, The
Netherlands}
\author[0000-0002-9542-3852]{Bastien Duboeuf}
\affiliation{University of Lyon, ENS de Lyon, CNRS, Laboratoire de Physique, F-69342, Lyon, France}
\author[0000-0001-6573-7773 ]{Samaya Nissanke}
\affiliation{GRAPPA, Anton Pannekoek Institute for Astronomy and Institute of High-Energy Physics, University of Amsterdam, Science Park 904, 1098 XH Amsterdam, The
Netherlands}
\affiliation{Nikhef, Science Park 105, 1098 XG Amsterdam, The Netherlands}
\author[0000-0003-4617-4738 ]{Francois Foucart}
\affiliation{Department of Physics \& Astronomy, University of New Hampshire, 9 Library Way, Durham NH 03824, USA}
\author[0000-0002-3394-6105]{Tanja Hinderer}
\affiliation{Institute for Theoretical Physics, Utrecht University, Princetonplein 5, 3584 CC, Utrecht, The Netherlands}
\author[0000-0002-7627-8688]{Andrew Williamson}
\affiliation{GRAPPA, Anton Pannekoek Institute for Astronomy and Institute of High-Energy Physics, University of Amsterdam, Science Park 904, 1098 XH Amsterdam, The
Netherlands}



\begin{abstract}

Detection of gravitational waves (GWs) from neutron star-black hole (NSBH) standard sirens can constrain the cosmological distance-redshift relationship and consequently provide local measurements of the Hubble constant ($H_0$), regardless of the detection of an electromagnetic (EM) counterpart: The presence of matter interaction terms in GWs breaks the degeneracy between the mass parameters and the redshift, allowing the simultaneous measurement of both the GW luminosity distance and redshift.
Although the tidally disrupted NSBH systems can have EM emission, the detection prospects of an EM counterpart will be limited to the local universe ($z < 0.8$ in the optical) in the era of the next generation GW detectors, Einstein Telescope (ET) and Cosmic Explorer (CE). 
However, the distinctive merger morphology, as well as the high redshift detectability of tidally-disrupted NSBH mergers
makes them promising standard siren candidates for this method.
Using recent constraints on the equation-of-state of NSs from multi-messenger observations of NICER and LIGO/Virgo/KAGRA, we show the prospects of measuring $H_{0}$ solely from GW observation of NSBH systems, achievable by ET and CE detectors.
We first analyse individual events to quantify the effect of the high-frequency ($\ge$ 500 Hz) tidal distortions on the inference of NS tidal deformability parameter ($\Lambda$) and hence on $H_0$. We find that disruptive mergers can constrain $\Lambda$ up to $\mathcal{O}(60\%)$  more precisely than non-disruptive ones. However, this precision is not sufficient to place stringent constraints on the $H_0$ parameter for individual events, due to the fundamental degeneracy between $H_{0}$ and the redshift.
By performing a Bayesian analysis on different sets of simulated NSBH data (up to $N=100$ events, corresponding to a timescale from several hours to a day observation)
in the ET+CE detector era, we find that NSBH systems could enable unbiased 4\% - 13\% precision on the estimate of $H_0$ (68\% credible interval).
This is a similar measurement precision found in studies analysing populations of NSBH mergers with EM counterparts in the LVKC O5 era ($\sim$ 2yrs).
\end{abstract}



\section{Introduction}
The value of the Hubble constant $H_{0}$, which quantifies the current expansion rate of the Universe, has been measured extensively since it was first established in 1929~\citep{Hubble168}. Even with the current high-precision measurements of $H_{0}$, the most recent local measurement $H_0 = 73.04 \pm 1.04\,\mathrm{km\, s^{-1}Mpc^{-1}}$ of the Hubble Space Telescope and Supernova H0 for the Equation of State (SH0eS) team~\citep{Riess:2021jrx}, highlights a level of $\approx 5\sigma$ tension with the constraint inferred by the Planck Collaboration $H_0 = 67.4 \pm 0.5\,\mathrm{km\,s^{-1}Mpc^{-1}}$~\citep{Planck:2018vyg}. Despite the ongoing efforts to find conclusive evidence of systematic errors in modeling the data of these experiments, or a compelling novel theoretical explanation, there is currently no agreement on the cause of the discrepancy in $H_{0}$ between the different measurements.

GW detection and sky-localization of merging binaries can provide a direct and independent local measurement of $H_{0}$, as first proposed by~\citet{1986Natur.323..310S}, and further analysed and advanced by~\cite{ Holz:2005df,PhysRevD.74.063006,Nissanke:2009kt,PhysRevD.85.023535,Nissanke:2013fka,PhysRevD.93.083511,Seto:2017swx,Chen:2017rfc,Fishbach:2018gjp,PhysRevLett.122.061105,PhysRevD.100.103523,Soares-Santos:2019irc,Palmese:2020aof,Mukherjee:2020hyn,Vasylyev:2020hgb,Chen:2020gek,Gayathri:2020fbl,Mukherjee:2020kki,PhysRevLett.121.021303,Feeney:2020kxk,DES:2019ccw,2021PhRvD.104f2009M,2020arXiv200702883B,2022MNRAS.512.1127G,2022MNRAS.511.2782C}. These systems are referred to as bright standard sirens, in case an EM follow up can be assigned to the event, and otherwise dark standard sirens.
The GW detection of the binary neutron star (BNS) system GW170817 and the electromagnetic (EM) identification of its host galaxy(~\cite{LIGOScientific:2017zic} and references therein) allowed the first application of the bright standard siren's approach, giving $H_{0}=70^{+12.0}_{-8.0}\,\mathrm{km\, s^{-1}Mpc^{-1}}$~\citep{Abbott:2017xzu}. This measurement was followed by improved estimate of $H_{0}=68.9 \pm 4.7\,\mathrm{km\, s^{-1}Mpc^{-1}}$ ~\citep{Hotokezaka:2018dfi}, using high angular resolution imaging of radio counterparts of GW170817, and later on estimated to $H_{0}=68.3^{+4.6}_{-4.5}\,\mathrm{km\, s^{-1}Mpc^{-1}}$~\cite{2021A&A...646A..65M} and $H_{0}=68.6^{+14.0}_{-8.5}\,\mathrm{km\, s^{-1}Mpc^{-1}}$ ~\citep{Nicolaou:2019cip}, by accounting for the systematic uncertainties that arise from the calculation of the peculiar velocity. 
Recently, the third gravitational wave catalogue was released, bringing the total number of GW detections to 90 events \citep{gwtc3}. Selecting 47 of these events, the Hubble constant was constrained to $H_{0}=68^{+13.0}_{-7.0}\,\mathrm{km\, s^{-1}Mpc^{-1}}$ when using the redshifted mass distribution, and $H_{0}=68^{+8.0}_{-6.0}\,\mathrm{km\, s^{-1}Mpc^{-1}}$ when combining the GW information with a galaxy catalog \citep{gwtc3_hubble}, and to $H_{0}=67^{+6.3}_{-3.8}\,\mathrm{km\, s^{-1}Mpc^{-1}}$ when using the GWTC-3 catalogue in combination with a galaxy catalogue~\citep{2022arXiv220303643M}.

The combined GW and EM detection of NSBH systems, however, is yet to be observed.
In general, NSBH systems are expected to be promising  standard sirens -- as both dark and bright candidates --since they have higher masses compared to BNSs, leading to mergers that occur at lower frequencies, potentially within the current and future ground-based detector bands, and also accessible at higher redshifts~\citep{Nissanke:2009kt,PhysRevLett.121.021303,Feeney:2020kxk}. 
The key difference between the NSBH systems and BNSs, which makes NSBH systems specifically interesting as dark standard sirens as well, is in their inspiral-merger phenomenology. 
For both systems, the tidally-induced deformations on the neutron star (NS) from the companion object (quantified by the NS tidal deformability parameter $\Lambda$) would increase the GW energy emission rate at the early inspiral stage in a similar manner -- for NSBH mergers this effect can increase even further if the BH is spinning fast.
Yet, closer to the merger of NSBHs, the strong tidal fields of the BH can, in some cases, significantly disrupt the NS, causing a sudden decrease in the GW amplitude at high frequencies and an accelerated merger, followed by mass ejection and formation of accretion torus around the BH and consequently, EM radiation. The other possible fate of the NS is that it plunges into the BH before getting highly disrupted and having a chance to emit any EM radiation. Qualitatively, whether the disruption happens or not, and how strong it is, depends primarily on the eccentricity, mass and spin of the BH, and the internal NS matter structure~\citep{Lattimer:1974slx,Vallisneri:1999nq,PhysRevD.86.124007,Deaton:2013sla,PhysRevD.92.081504,Foucart:2020ats,PhysRevD.79.044030,Etienne:2008re,Foucart:2012vn,Stephens:2011as}. 

In the highly-disruptive cases that lead to EM radiation, NSBH systems can be used as bright standard sirens~\citep{Feeney:2020kxk,Vitale:2018wlg,Nissanke:2009kt} with the spectroscopic redshift for the host being obtained with very high accuracy. Identifying such EM counterparts, however, remains challenging as the current and future planned EM facilities have limitations in the sky coverage~\citep{2012ApJ...746...48M,2013ApJ...767..124N,Raaijmakers:2021slr,Chase22,2019BAAS...51c.276S} and hence many of such events will be too far ($z>1$) to be detectable by wide field optical and radio telescopes.
Moreover, the probability of
detecting an EM counterpart could strongly depend on the orientation angle of the system.

In the cases when no EM counterpart is observed or generated, GW measurements of BNS or NSBH mergers can solely constrain the distance-redshift relation (hence the cosmological parameters), with a method first proposed by~\citet{PhysRevLett.108.091101} and applied to BNS mergers more recently by ~\citet{PhysRevD.95.043502, Chatterjee:2021xrm} and \citet{2022arXiv220311756G} (see~\cite{2019ApJ...883L..42F} for a technique applicable to BBHs). Tidal deformation in a binary system affects the transfer of GW energy and consequently, the GW phase and amplitude evolution and also accelerates the coalescence. In the absence of such matter effects, the mass parameters of GW signals are degenerate with the redshift $z$, resulting in the detection of redshifted masses $m_{NS,d}$ measured by the detector.
However, the tidal corrections in the GW signals depend on the physical masses \textit{i.e.}, the source-frame NS masses such that $m_{NS,s}=m_{NS,d}/(1+z)$. Therefore, these $\Lambda$-dependent corrections break the degeneracies in the waveform and thus allow the simultaneous estimation of the GW luminosity distance $d_L$ and redshift $z$, from the waveform's amplitude and phase respectively. 


To constrain $H_{0}$ 
with this approach, we need independently derived information on the NS matter effects, either by assuming a known NS equation-of-states (EoS) \citep{PhysRevLett.108.091101} or by using some form of parameterization -- such as Taylor expansion of $\Lambda$ in terms of NS mass -- in an EoS-insensitive way. An example of latter is using the so-called universal binary-Love relations~\citep{Yagi:2016bkt,Doneva:2017jop}, which fits $\Lambda$ around a fiducial NS mass value for which the tidal parameter is known. This approach is so far only applied to GW170817 for which the NS has a fiducial mass that lies in the steepest region of mass-radius plot. Hence, probing the extreme cases and also very high NS masses with this approach can be limited.\\
In this paper, we use a new approach for modelling the viable EoS parameters. We use the posterior EoS samples from \citet{Raaijmakers2021a}, which are inferred from a combination of multimessenger astrophysical observations and low-density nuclear calculations done within a chiral effective field theory framework \citep{Hebeler13}. The astrophysical observations include the two mass-radius measurements of millisecond pulsars PSR J0030+0451 \citep{MillerJ0030,RileyJ0030} and PSR J0740+6620 \citep{MillerJ0740, RileyJ0740} by NASA's Neutron Star Interior Composition Explorer (NICER)~\citep{2016SPIE.9905E..1HG} and the tidal deformability measurement from GW170817 and its accompanying EM counterpart AT2017gfo, and the low signal-to-noise ratio (SNR) event GW190425 \citep[see also][]{GuerraChaves:2019foa,Pang:2021jta,Dietrich:2020efo} for which no EM counterpart was observed.

In addition to the uncertainties in the modelling techniques, the significance with which $H_{0}$ can be inferred also depends on how well the tidal deformability parameter $\Lambda$ can be constrained by the observed GWs. Note that measuring $H_{0}$ with this approach is not applicable to the recently detected NSBH systems GW200105 and GW200115~\citep{LIGOScientific:2021qlt} due to the low SNR of their detected signals and the uninformative constraints on their $\Lambda$ parameter with the current detectors. This itself is partly due to the lack of information from their merger stage and partially due to the fact that they are (very likely) non-disrupting systems which can make them uninformative even at high SNRs.

\begin{figure*}
\includegraphics[width=0.9\linewidth]{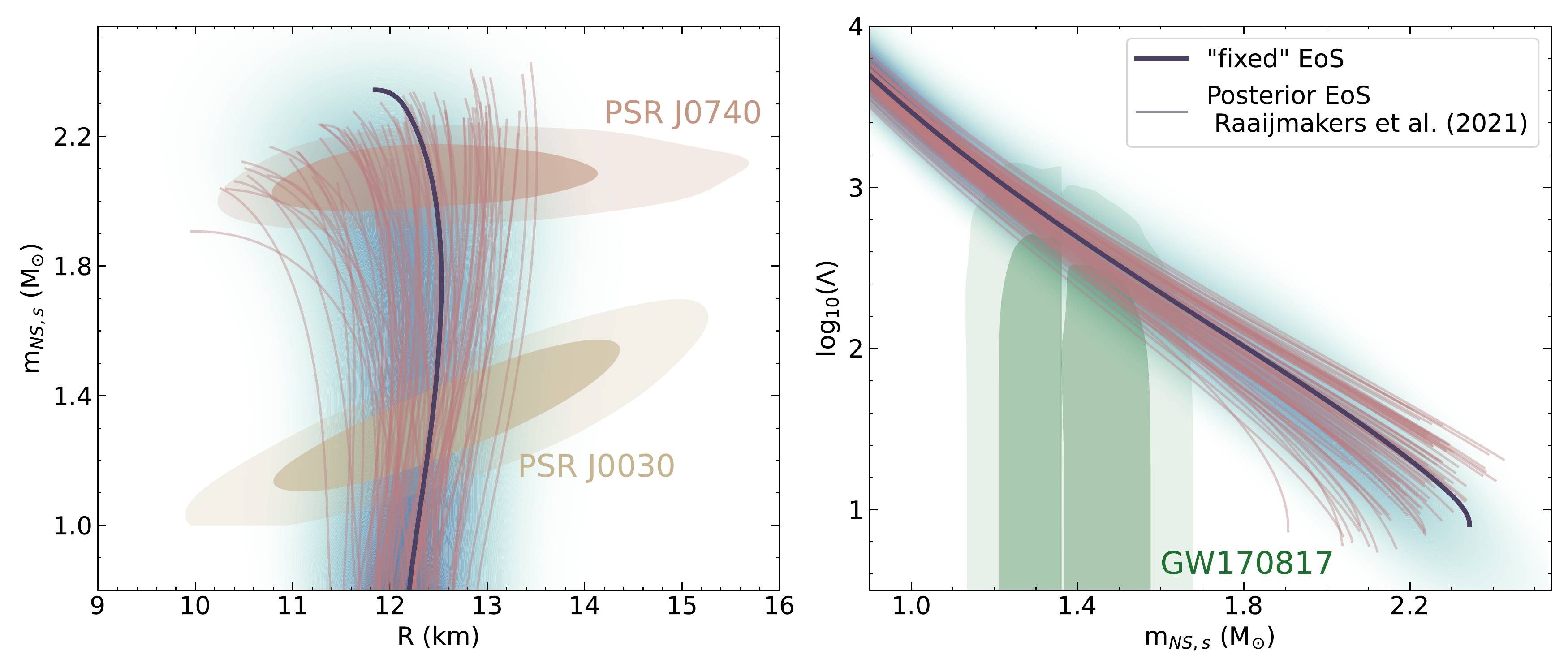}
\centering
\caption{The combined mass-radius constraints found by LIGO/Virgo and NICER observations, together with examples of possible viable EoS relations (pink curves) based on~\citet{Raaijmakers2021a}. The black line shows the EoS with the highest posterior support (the blue shaded region indicates the posterior distribution) and is the chosen EoS relation for generating initial GW signals. The green regions on the right plot show the multi-messenger constraints found from the \textbf{}NS event GW170817.}
\label{fig:eos}
\end{figure*}

The effect of high-frequency tidal disruptions on the GW amplitude of NSBH systems is usually modeled by introducing a cut-off frequency parameter which roughly indicates whether a merger is disruptive or not, and also marks the beginning of a possible disruption~\citep{Vallisneri:1999nq,Ferrari:2008nr,PhysRevD.81.064026,PhysRevD.89.043009,Pannarale:2015jka,PhysRevD.92.081504,PhysRevD.89.043009}. This is a distinctive feature of the waveform emitted by NSBH binaries: If the disruption, characterised by the cut-off frequency, takes place before the NS crosses the BH's innermost-stable-circular orbit (ISCO), then the ejected material from the NS can form a disk around the BH and suppress the amplitude of the waveform, leading to an accelerated merger with no post-merger GW signal.
On the other hand, in the case of a non-disruptive merger (\textit{e.g.}, when the BH mass is very large), the waveform is shown to be comparable to binary BH waveforms~\citep{PhysRevD.84.104017}, where, instead, the high-frequency amplitude is governed by the ringdown of the companion BH. Therefore, the main difference between a BBH and a NSBH signal that is non-disruptive, is mainly in their GW phasing behaviour. 

In this paper we determine constraints on $\Lambda$ -- and consequently on $H_0$ -- by performing Bayesian parameter estimation on simulated NSBH binary systems in the next generation detectors, ET~\citep{Maggiore:2019uih} +CE detector era~\citep{Evans:2021gyd,Reitze:2019iox}, using the previously mentioned multi-messenger constraints on NS EoS to model $\Lambda$. We compare the bounds on $\Lambda$ and $H_{0}$ derived from the tidally disrupted and non-disrupted mergers to see how the high-frequency NS disruption effects in the waveforms can improve the overall parameter inference in the era of these next generation observatories. Our primary interest is in the tidally-disrupted systems (\textit{i.e.}, systems with low mass-ratios and possibly high prograde BH spins) as most of them can merge inside the ET+CEs detector bandwidth while the current ground-based detectors can not capture their mergers.

In order to model the GW strain data, we use the NSBH specific GW waveform model \texttt{IMRPhenomNSBH} ~\citep{Thompson:2020nei} (hereafter referred to as Phenom-NSBH).
We then analyse the prospects of measuring $H_{0}$ by performing a two-step Bayesian parameter estimation on simulated individual NSBH systems, as well as catalogues of NSBH events with different number of simulated events. We use the same waveform model for both the injection and recovery of the waveforms.

The paper is organized as follows. In Sec.~\ref{sec:methods} we introduce our approach for modelling the tidal deformability parameter, and give an overview of the  tidal modelling in the chosen waveform model. After explaining the details of source and population simulations in Sec.~\ref{sec:simulation}, we describe the details of the Bayesian statistical framework used in our analysis in Sec.~\ref{sec:gwinference}. We present the results of analysing single NSBH systems and stacked catalogs in Sec.~\ref{sec:results}. Sec.~\ref{sec:conclustions} summaries our conclusions.

\section{Methods}\label{sec:methods}
\subsection{Tidal deformability model}
\label{subsec:tidalmodel}
During the evolution of a NSBH binary, the tidal fields of the BH produce deformations in the companion NS. These deformations depend on the NS matter properties, predominantly through an EoS-dependent dimensionless tidal deformability parameter $\Lambda$ defined as:
\begin{eqnarray}\label{Lambda}
    \Lambda=\frac{2}{3}\frac{k_2}{\mathcal{C}^5}=\frac{2}{3} k_2 \left(\frac{c^2 R}{Gm_{NS,s}}\right)^5=\lambda G \left(\frac{c^2 (1+z)}{Gm_{NS,d}}\right)^5
    \,,
\end{eqnarray}
where $m_{NS,s}$ is the source-frame mass of the NS, $\mathcal{C}=Gm_{NS,s}/R c^2$ is the NS compactness, and $k_2$ is the dimensionless relativistic quadrupole tidal love number such that $\lambda=2/3 R^{5} k_{2}G^{-1}$ (with $R$ being the NS radius) characterizes the strength of the induced quadrupole given an external tidal field \citep{Hinderer:2009ca, Flanagan:1997fn}. The right hand side of the equation shows the redshift dependency once we transform to detector-frame mass $m_{NS,d}=m_{NS,s}(1+z)$.

We base our modeling of the tidal deformability parameter on the EoS constraints inferred by \cite{Raaijmakers2021a} \citep[For other multimessenger EoS constraints, see e.g.][]{Dietrich:2020efo, Almamun21, Huth22}. In this work, the EoS is constrained by employing a parameterized high-density EoS, coupled to low-density NS matter calculations within a chiral effective field theory framework \citep{Hebeler13}. Posterior distributions on the EoS parameters are then obtained by combining information from NICER's mass and radius measurements of the pulsars PSR J0030+0451 \citep{MillerJ0030, RileyJ0030} and PSR J0740+6620 \citep{MillerJ0740, RileyJ0740}, and measurements of the tidal deformability from GW170817 (together with its optical counterpart AT2017gfo) and GW190425. 
Figure~\ref{fig:eos} shows the mass-radius and mass-tidal deformability constraints as found by the NICER and LIGO/Virgo observations as well as the possible EoS relations consistent with the posterior distribution found in \citet{Raaijmakers2021a}. 

\begin{figure*}
\includegraphics[width=0.7\linewidth]{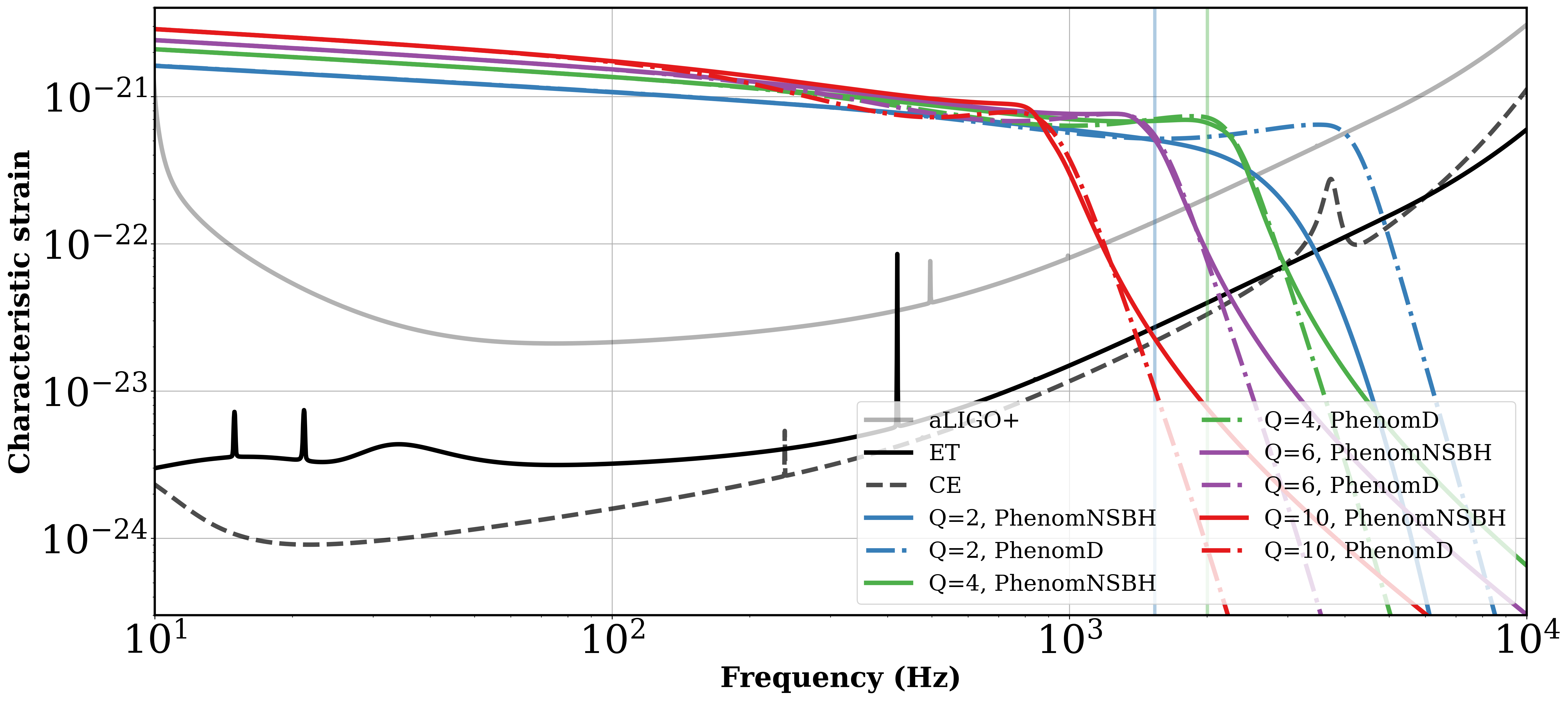}
\centering
\caption{The dimensionless sensitivity curves $\sqrt{S_{n}\times f}$ and the characteristic signals $h_c=f\times \tilde{h}(f)$ as a function of frequency $f$. The solid lines correspond to the Phenom-NSBH
model and the dot-dashed lines to the corresponding BBH binary modeled signal IMRPhenomD for a system at $z=0.02$ ($d_L \approx 100\, \mathrm{Mpc}$).
As the mass-ratio of the system increases, the mergers change from total disruption of the NS, with amplitude of the waveform being exponentially suppressed at high frequencies, to non-disruptive NS for which the waveform is comparable to the BBH waveform. The shaded blue and green lines correspond to cut-off frequencies $f_{cut}(Q=2)= 1535$\,Hz and $f_{cut}(Q=4)= 2000$\,Hz.}
\label{fig:strain}
\end{figure*}



Instead of sampling $\Lambda$, we sample EoSs from the posterior distribution of \citet{Raaijmakers2021a}. For a drawn EoS, a value of $\Lambda$ can then be assigned based on a given $m_{NS,s}$. By considering a broad set of EoSs as a prior in our analysis, we can take into account the uncertainties in the NS's microphysics. However, we remain dependent on our choice of EoS parameterization, which may introduce a systematic bias. We specifically take the results from the piecewise-polytropic parameterization employed in \citet{Raaijmakers2021a}, but note that many other high-density parameterizations exist \citep[][]{Lindblom18,Greif19,Capano:2019eae, Boyle20} as well as non-parametric methods \citep[][]{Landry19, Essick:2019ldf, Landry20, LegredJ0740}. Although this approach does not provide an EoS-insensitive model, the advantage of this approach is that it does not require an approximate analytical modeling of the $\Lambda$ parameter and is based on the current empirical multi-messenger constraints.
\subsection{Waveform model}
To model the GW waveform, we use the Phenom-NSBH model~\citep{Thompson:2020nei}, which extends the analytical inspiral wave to incorporate the merger and post-merger dynamics through calibration with numerical relativity simulated waveforms. This waveform is built on IMRPhenomC model for the amplitude and IMRPhenomD model for the phasing, covering non-precessing systems with mass ratios ($Q= m_{BH}/m_{NS}$) from equal-mass up to 15, BH spins aligned with the orbital angular momentum up to a dimensionless value of $|\chi_{BH}|= 0.5$, and $\Lambda$ ranging from 0 (the BBH limit) to 5000. Note however that, in the case of $\chi_{BH}<0$, the model amplitude is calibrated to NR simulations up to $Q<4$ and hence may not perform well beyond this mass-ratio.
\footnote{An alternative state of the art model for NSBH mergers is based on the effective-one-body approximation; see~\cite{Matas:2020wab}.}
In modelling the Phenom-NSBH waveform, the tidal corrections to the GW phase are incorporated as post-Newtonian (PN) spin-induced quadrupole corrections at 5\,PN and 6\,PN order~\citep{PhysRevD.81.123016,PhysRevD.83.084051}. To first order, the change in GW phase scales linearly with $\Lambda$, through a dimensionless quantity defined as:
\begin{equation}
   \tilde{\Lambda} = \frac{48}{39}\frac{m_{NS,s}^4 (m_{NS,s}+12 m_{BH,s})}{(m_{NS,s}+m_{BH,s})^5}\Lambda\,.
\end{equation}
This scaling shows that finite size effects are expected to be mainly detectable for NSBHs involving low mass BHs. As the mass ratio increases,
tidal effects scale away as $Q^{-4}$ in the phase, making the non-disruptive NSBH signal hard to differentiate from a BBH signal. Therefore, in the case of short ($\approx$ less than 100 s) or low SNR ($\approx$ less than 30) signals, the only differences between these two waveform models -- non-disruptive NSBH systems and BBH systems-- could be the slightly different properties of the remnant quantities after merger which are hard to distinguish with the current GW detectors~\citep{Foucart:2013psa,2014PhRvL.113i1104T}. Note that, in the case of possibly long and loud GW signals, the accumulated waveform phase difference between a BBH and a disruptive NSBH system (due to the presence of tidal terms in the latter case) can lead to bounds on $\Lambda$ such that the BBH case ($\Lambda=0$) gets excluded. However the overall bounds on $\Lambda$ for the non-disruptive systems are still expected to be broader than the disruptive ones, and generally uninformative in most of the cases.

For the GW amplitude in the PhenomNSBH model, a semi-analytical modelling of tidal effects at the late-inspiral is adapted ~\citep{Pannarale:2015jka}. In this modelling, the merger of a NSBH binary is considered disruptive whenever the mass ratio $Q<Q_{D}(\mathcal{C},\chi)$, with the threshold being fitted by:
\begin{equation}
    Q_{D} = \sum_{i,j=0}^3 a_{i,j}\mathcal{C}^i \chi^j , \hspace{0.5cm} i+j \leq 3,
\end{equation}
where $\chi$ is the BH spin parameter and the fitting parameters $a_{i,j}$ are as given in~\cite{PhysRevD.92.081504}.
The corresponding fitted cut-off frequency $f_{cut}$ to this threshold is given by:
\begin{equation}
   f_{cut}=\sum_{i,j,k=0}^{3}f_{ijk}\mathcal{C}^{i}Q^{j}\chi^{k}, \hspace{0.5cm} i+j+k \leq 3\,,
\end{equation}
With the fitting parameters $f_{ijk}$~\citep{PhysRevD.92.081504}.
The fitted threshold parameters depend on the EoS only implicitly and through the NS compactness.
Moreover, the aforementioned waveform model approximates $\mathcal{C}$ in terms of $\Lambda$ by using the quasi-universal relation on the compactness~\citep{Yagi:2016bkt}:
\begin{equation}\label{YYcompactness}
  \mathcal{C}=0.371-0.0391\,\mathrm{log}(\Lambda_{NS})+0.001056\,\mathrm{log}(\Lambda_{NS})^2\,.
\end{equation}
This allows us to quantify the disruptions of each merger solely based on the waveform parameters that can be derived from detected GW data.

Figure~\ref{fig:strain} shows non-spinning NSBH GW characteristic signal $h_c = f \times \tilde{h}(f)$ at $z=0.02$ ($\approx 100\,\mathrm{Mpc}$), for disruptive and non-disruptive mergers as compared to their BBH analogue signal. For the disruptive mergers, the $f_{cut}$ is also indicated and shows the approximate frequency at which the waveforms start to deviate from a BBH signal. As expected, increasing the mass ratio would change the mergers from being totally disruptive to non-disruptive. Although not shown here, we have also investigated that disruptive waveforms with a higher $\Lambda$ have larger deviations from their BBH counterpart waveforms.\footnote{Note that the high frequency amplitude increase of the BBH waveforms for the $Q=4,6,10$ cases, as well as the mismatch with the dip in the BBH waveform, are due to the unphysical artefact in the waveform model and do not represent physical features (see also Fig.~1 of ~\citep{Thompson:2020nei}). }

\subsection{Simulations and sources}\label{sec:simulation}
We simulate samples of individual NSBH binary events in the detector bands of the 3-interferometer ET and CE (i.e., 5\,Hz to 4\,kHz), and using a sampling frequency rate of 4096 Hz.  
For the possible CE sites, we choose the Northwestern USA and Southeastern Australia location (see \citet{Gossan:2021eqe} for the detector coordinate details and further work). The ET site implemented in \texttt{lalsuite}~\citep{lalsuite}, and hence used in this work, is same as the location of the Virgo detector.
Having multiple detectors allows us to localize the GW sources and infer the different GW polarization content, thus allowing one to break the degeneracy between the luminosity distance and inclination angle for high SNR systems~\citep{2022arXiv220211048B}. Note that we are assuming that the calibration errors will be less than $1\%$ in the amplitude and phasing (see~\citet{2022arXiv220403614H} for more detailed analysis).

For the injected cosmological parameters we use $H_{0}=67.4\, \mathrm{km\,s^{-1} Mpc^{-1}}$, along with the rest of the parameters as reported in~\citet{Planck:2018vyg}. In order to isolate the effect of tidal interactions on the inference, as well as due to the limitations of the waveform model, we do not consider the effects of spin precession and orbital eccentricity here. The inclusion of spin precession, however, has been shown to improve the $H_0$ inference by breaking the distance-inclination degeneracy, in some NSBH systems~(\textit{cf.}~\citet{Vitale:2018wlg}).

We generate two sets of simulations:\\
\textbf{I)} Set of individual mock binary samples, that will allow us to compare the disruptive and non-disruptive mergers and their effect on the $H_0$ inference. We generate such mock binary samples, having  fixed $m_{NS,d}=1.4 M_{\odot}, \Lambda= 490$. This choice of $\Lambda$ corresponds to fixing the EoS parameter to the maximum-a-posteriori relation based on the results of \cite{Raaijmakers2021a}, inferred from NICER's pulsar measurements and the BNS GW events detected by LIGO, Virgo and KAGRA (see also section \ref{subsec:tidalmodel} and the black line of Fig.~\ref{fig:eos}). We vary the mass ratios to be $Q=m_{BH,d}/m_{NS,d}=\{2,4,6\}$ , at a certain sky position and polarization of (ra, dec, polarization) = $(3.45,\, -0.41,\, 2.35)$ rad. In order to study the impact of GW parameters on the inference, we also choose the BH spins to be $\chi_{BH}=\{0, 0.5, -0.5\}$ and consider systems with different inclination angles $\theta_{jn}$ bracketing to the two extreme cases of edge-on and face-on binaries. Increasing the distance, as expected, would broaden the inferred bounds on $\Lambda$, as the disruptive binaries would merge outside the detector's sensitivity band. Here we consider binaries located at $z = 0.07$ and $z=0.2$.\\ 
\textbf{II)} Due to possible degeneracies between parameters and due to measurement uncertainties, the analysis of such single events may lead to  multi-modal $H_0$ measurements. To overcome these limitations, we also analyse stacked catalog of mock binary samples of size $N$=10, 70 and 100, in the ET+CEs detector era.  We distribute the binary systems uniformly in sky location, orientation, and volume ($\propto d_{L}^2$). We also sample from uniform distributions of BH and NS source masses and spins, same as that of Sec.~\ref{sec:gwinference}. For generating these samples, we also vary the initial choice of EoS parameter by uniformly sampling $\approx 3000$ EoS choices based on the posterior distribution of ~\citet{Raaijmakers2021a} . We consider the systems with the network SNR of above $8$ to be detectable. 

\begin{figure*}
\includegraphics[width=0.85\linewidth]{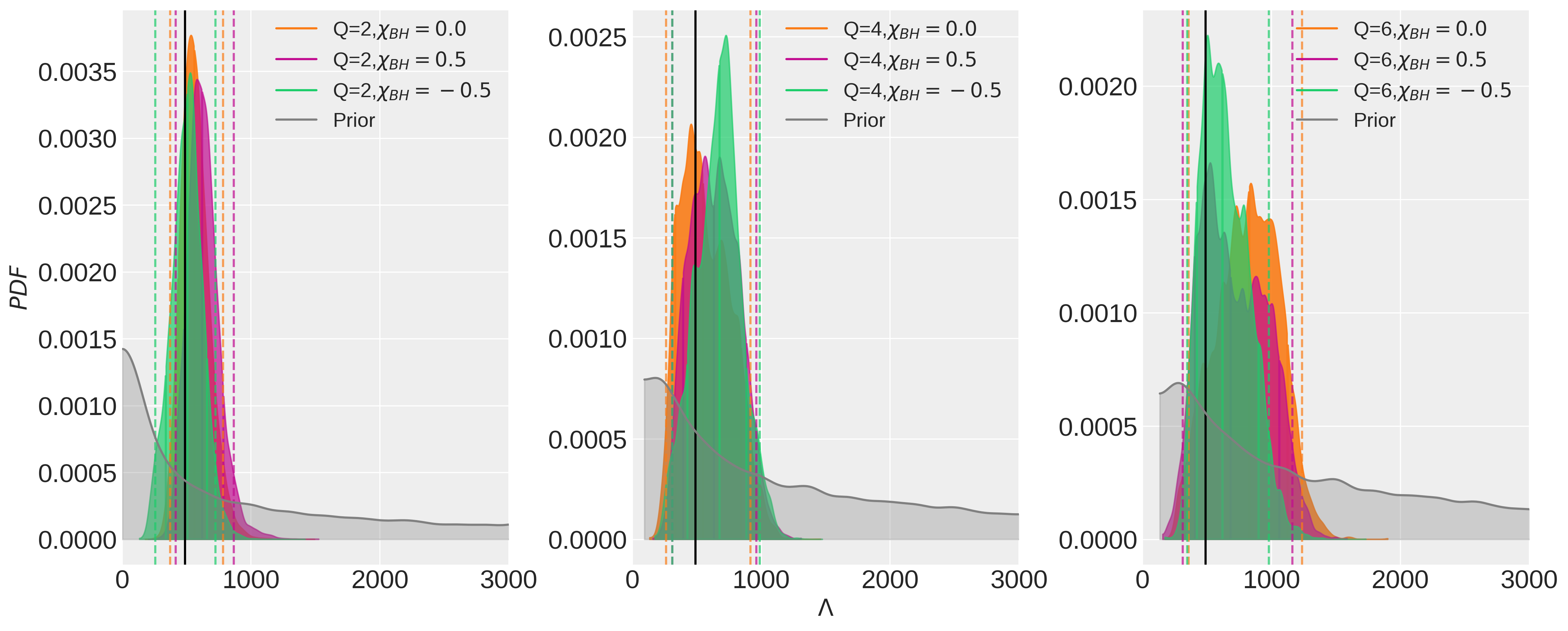}
\includegraphics[width=0.85\linewidth]{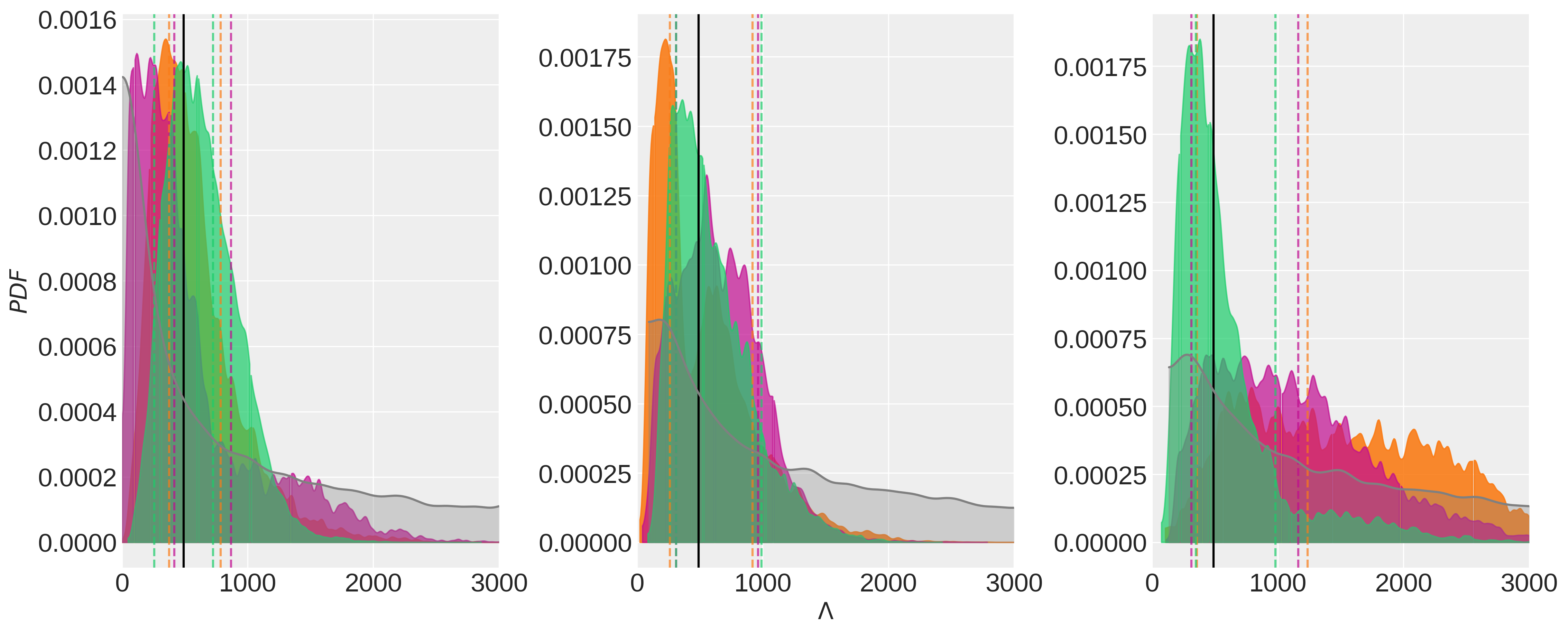}
\centering
\caption{The posterior probability density and the 90\% credible regions of $\Lambda$ parameter as measured for $Q=2$\,(disruptive),$Q=4$\,(mildly-disruptive), and $Q=6$\,(non-disruptive) systems located at $z=0.07$ (top) and $z=0.2$ (bottom). The orange, pink and green shaded regions correspond to the choice of $\chi_{BH}=0$, $\chi_{BH}=0.5$ and $\chi_{BH}=-0.5$, respectively. The black vertical lines correspond to the injected value of $\Lambda=490$ and the gray shaded regions show the implied prior on $\Lambda$ parameter.}
\label{fig:Lambda}
\end{figure*}
\subsection{Statistical framework}
\subsubsection{Gravitational wave parameter inference}\label{sec:gwinference}
In order to perform a probabilistic inference of $\Lambda$ and subsequently $H_{0}$ through Bayesian analysis (as described in, for example, Appendix\,B of~\cite{PhysRevX.9.031040}), we first evaluate the GW probability distribution function (PDF) $P(\theta_{i}|\mathbf{x})$ with $\mathbf{x}$ being the simulated GW strain observation and $\theta_{i}$ the set of waveform parameters that we want to estimate. 
For this, we marginalize the GW likelihood over the coalescence phase.
We sample the marginal GW likelihood using the relative binning approach~\citep{Zackay:2018qdy}, as implemented in the inference library \texttt{PyCBC}~\citep{Biwer:2018osg} together with the \texttt{DYNESTY} nested-sampler~\citep{10.1093/mnras/staa278}, using 2100-2500 live points. The relative binning method allows for fast analysis of GW signals by assuming that the difference between adequate waveforms in the frequency domain is describable by a smoothly varying perturbation. Having the fiducial gravitational waveforms
close to where the likelihood peaks, this approach reduces the number of frequency points for the evaluation of the waveforms to $\mathcal{O}(10^2)$, as compared to $\mathcal{O}(10^7)$ for traditional GW parameter inference.

The prior choice on each parameter is as follows. We draw uniformly distributed BH masses from $\mathrm{P}(m_{BH,s})=\mathrm{U}(2.5 M_{\odot}, 12 M_{\odot})$ 
with the lower-limit set in such a way to allow BHs in the mass-gap region and the upper limit in such a way to avoid running into high values of $Q$ for which the waveform approximation does not function.
The NS masses are drawn from $\mathrm{P}(m_{NS,s})=\mathrm{U}(1 M_{\odot}, 2.5 M_{\odot})$, with the upper limit chosen to be above the current estimates of the maximum mass of NSs \citep{LegredJ0740, Pang0740}. We consider aligned-spin binary components with the dimensionless BH and NS spin parameter drawn from  $\mathrm{P}(\chi_{BH})=\mathrm{U}(-0.5, 0.5)$ and low-spin prior of $\mathrm{P}(\chi_{NS})=\mathrm{U}(-0.05, 0.05)$, respectively. In a future work, we will examine astrophysically motivated populations for NSBH mergers~\citep{Broekgaarden21,Boersma:2022yoz}.
The redshift parameter is sampled from $\mathrm{P}(z)=\mathrm{U}(0, 1.5)$.
The inclination angle, sky position, phase and polarization angles are isotropically distributed. The luminosity distance $d_{L}$ is distributed uniformly in the comoving volume such that $\mathrm{P}(d_{L})\propto d_L^2$.
For the EoS parameter, we choose a prior based on the multi-messenger constraints ( as explained in Sec.~\ref{subsec:tidalmodel}), covering up to $3000$ viable EoS choices. Consequently, for each realisation of the sampler, the $\Lambda$ parameter is derived based on the sampled set of $\{z,\, m_{NS,d}\, \textrm{EoS}\}$ parameters, using the pycbc implemented transformation function \texttt{Lambda\_from\_multiple\_tov\_files}. For this we require the sampling to be done in terms of $z$ and $m_{NS,d}$ instead of $\Lambda$ and $m_{NS,d}$.  Moreover, in order to speed-up the inference, we sample in detector-frame chirp mass and mass ratio instead of the individual masses.

\subsubsection{Hubble constant inference}\label{H0Inf}
We estimate $H_{0}$ using a two-step Bayesian inference analysis.
In order to estimate $H_{0}$ from the GW data, we use the redshift-distance relation as given by:
\begin{equation}\label{dl-z}
    d_{L}=\frac{c(1+z)}{H_{0}}\int_{0}^{z}\frac{dz'}{E(z')}\,,
\end{equation}
where $E(z')=\sqrt{\Omega_{r}(1+z')^4+\Omega_m(1+z')^3+\Omega_{DE}}$ corresponds to the assumption of a flat universe and $\Omega_{r},\Omega_{m},\Omega_{DE}$ are the radiation, matter and dark energy energy densities respectively. In this analysis we use the third order Taylor expansion of Eq.~\ref{dl-z} around $z=0$, i.e. for low redshifts.

Having the GW strain data $\mathbf{x}$ of a single event, the posterior on $H_0$ can be obtained from the semi- marginalized GW likelihood and the prior $P_0(H_0)$:
\begin{equation}\label{H0PDF}
    \frac{P(H_{0}|\mathbf{x})}{ P_0(H_0)}=
    \int P(\mathbf{x},\Lambda,d_L,m_{NS,d}|H_{0})\,dd_L\, d\Lambda\, dm_{NS,d}
    \,\,,
\end{equation}
where $P(\mathbf{x},\Lambda,d_L,m_{NS,d}|H_{0})$ is the GW likelihood marginalised over all the parameters other than $\{d_L,\Lambda, m_{NS,d} \}$. Equations eq.~\ref{Lambda} and eq.~\ref{dl-z} allow us to write down $d_{L}$ as a function of cosmological parameters and the GW inferred parameters $\{\Lambda ,\,m_{NS,d}\}$ such that $d_{L}=d_{L}(\Lambda ,\,m_{NS,d} ,\,H_{0})$. Therefore the marginalized likelihood can be further expanded as:
\begin{eqnarray}
    &&\int P(\mathbf{x},\Lambda,d_L,m_{NS,d}|H_{0})dd_L d\Lambda dm_{NS,d}\nonumber\\
    &&=\int P(\mathbf{x}|\Lambda,m_{NS,d},d_L)P(d_{L}|m_{NS,d},\Lambda,H_0)\nonumber\\
    &&\quad\times P_{0}(\Lambda, m_{NS,d},d_L|H_0)
    \,dd_L\, d\Lambda\, dm_{NS,d}\,,
\end{eqnarray}
where again, $P_0$ shows the prior on each parameter. The constraint between parameters is defined through $P(d_{L}|m_{NS,d},\Lambda,H_0)$ and can be replaced by a delta function such that:
\begin{eqnarray}
&& \int P(\mathbf{x},\Lambda,d_L,m_{NS,d}|H_{0})dd_L\, d\Lambda \,dm_{NS}\nonumber\\
    &&=\int P(\mathbf{x}|\Lambda,m_{NS,d},d_L)\delta(d_L-\hat{d}_L[ m_{NS,d},\Lambda,H_0])\nonumber\\
    &&\quad\times P_{0}(\Lambda, m_{NS,d},d_L|H_0)
    \,
    dd_L\, d\Lambda\,dm_{NS,d}\nonumber\\
    &&= P(\mathbf{x}|\hat{d}_{L}[m_{NS,d}\,\Lambda,H_{0}])P_0(H_0) =\mathcal{L}(H_0)
    \,,
\end{eqnarray}
where in the last line we have applied the delta function and also marginalized over the rest of the parameters.

We perform the sampling of the semi-marginal likelihood $\mathcal{L}(H_0)=P(\mathbf{x}|\hat{d}_{L}[m_{NS,d},\Lambda,H_{0}])$ using \texttt{pymultinest}~\citep{2014A&A...564A.125B}. In order to do so, we first perform a kernel density estimation fit to the semi-marginal likelihood using \texttt{kalepy}~\citep{Kelley2021}.
We assume a flat prior on $H_{0}$ of $\mathrm{P}_{0}(H_{0})=\mathrm{U}(10,300)\mathrm{\,km\, s^{-1}Mpc^{-1}}$.
In addition to this method, we have also performed the direct sampling of $H_0$ through \texttt{pycbc} and recover similar results, yet the two-step inference shows better convergence in some cases such as for low-redshift systems.
\section{Results}\label{sec:results}
\subsection{Single events: $\Lambda$ detectability}\label{sec:singlelambda}

It is widely expected that disruptive NSBH mergers, once visible in a detector band, will allow for a more accurate measurement of the NS tidal effects, and hence the redshift. In order to quantify the effect of high-frequency tidal disruption on the parameter inference, we analyse a selected sample of single NSBH events with different $Q$ and $\chi_{BH}$. During the inference, the EoS parameter is being varied freely, with a prior consistent with the posterior distribution found in~\cite{Raaijmakers2021a} (see Sec.~\ref{fig:Lambda}).

The derived constraints on the $\Lambda$ parameter are shown in Fig.~\ref{fig:Lambda}, for systems with $z=0.07$ (top, $d_L \approx 300\,\mathrm{Mpc}$) and $z=0.2$ (bottom, $d_L\approx 1\, \mathrm{Gpc}$). All the systems that are located at $z=0.07$ merge within the ET+CEs detectors bandwidth. The network SNRs for the top panel are $\approx 367, 470, 541$, and for the bottom panel are $\approx 107, 138,159$ for the $Q=2,4,6$, respectively. The initial inclination angle is chosen as $\theta_{ij}= 90^{\circ}$ for these plots. Due to the high SNR of the selected events, we have shown that the results remain consistent once changing $\theta_{ij}$. This is not generally true for systems at lower SNR and there we can clearly see the effect of $\theta_{ij}$ on the inference of $d_L$, and consequently $H_0$.

Focusing on the non-spinning cases considered here, we see that, the disruptive-mergers ($Q=2$) constrain $\Lambda$ ($95\%$ credible interval) to $\Lambda=570^{\,+216}_{\,-195}$ (top) and $\Lambda = 578^{\,+635}_{\,-516}$ (bottom), with the relative error of $\approx 16\%$ in both cases. In the case of highly-disruptive mergers ($Q=6$), these values worsen and are given as $\Lambda=868^{\,+420}_{\,-460}$ (top) and $\Lambda = 1549^{\,+1284}_{\,-1223}$ (bottom), with the relative error of $\approx 77\%$ in the former case, and an uninformative constraint on the latter case.
This analysis does not clearly show the expected positive (negative) effect of prograde (retrograde) BH spin on the inference of $\Lambda$. Possible sources of limitations can be the degeneracies (such as the degeneracy between the reduced mass, entering at 1\,PN order, and the spin, entering at 1.5\,PN order~\citep{1994PhRvD..49.2658C}), as well as the limitation of Phenom\_NSBH model at prograde spins: we anticipate that a more detailed analysis of the BH spin effects on $\Lambda$ inference is not feasible with the current limitations of the GW waveform models.

In general, having a flat prior on $\Lambda$, exclusion of $\Lambda=0$ value from the posterior samples would suggest that GW information alone can distinguish NSBH systems from a BBH merger. This may happen if a merger happens in-band for a detector system, or by having a long detected signal duration for an event: over a long inspiral time, the BBH and non-disruptive NSBH signals can get a dephasing of $\approx \mathcal{O}(1)$ radians from the point-particle tidal effects, which makes the waveforms non-identical. However, it is important to keep in mind that, in our analysis, the inferred bounds also depend on the $\{m_{NS},z, \mathrm{EoS}\}$ priors which may apriori exclude the $\Lambda=0$ value. More specifically, the \texttt{Lambda\_from\_multiple\_tov\_files} function assigns a $\Lambda=0$ value to each sampled $m_{NS,s}$ that exceeds the maximum allowed NS physical mass for a given EoS parameter file, and hence allows for some of the systems to be predicted as BBH systems. This implied prior on $\Lambda$ is shown in the gray shaded regions of Fig.~\ref{fig:Lambda}. For the middle and right panels, the minimum allowed $\Lambda$ values of the prior are $91$ and $132$, respectively.
In the cases where $\Lambda=0$ is allowed by the implied prior on $\Lambda$ (left panels), exclusion of this value from the posterior samples would suggest that GW information alone can distinguish this system from a BBH merger, which is critical for the validity of this approach on real GW data. This can be seen in the top left panel of Fig.~\ref{fig:Lambda} but not in the bottom left panel, due to the lower SNRs of the latter systems.

\subsection{Single events: pair plots and statistical uncertainties}
\begin{figure*}
\includegraphics[width=0.9\linewidth]{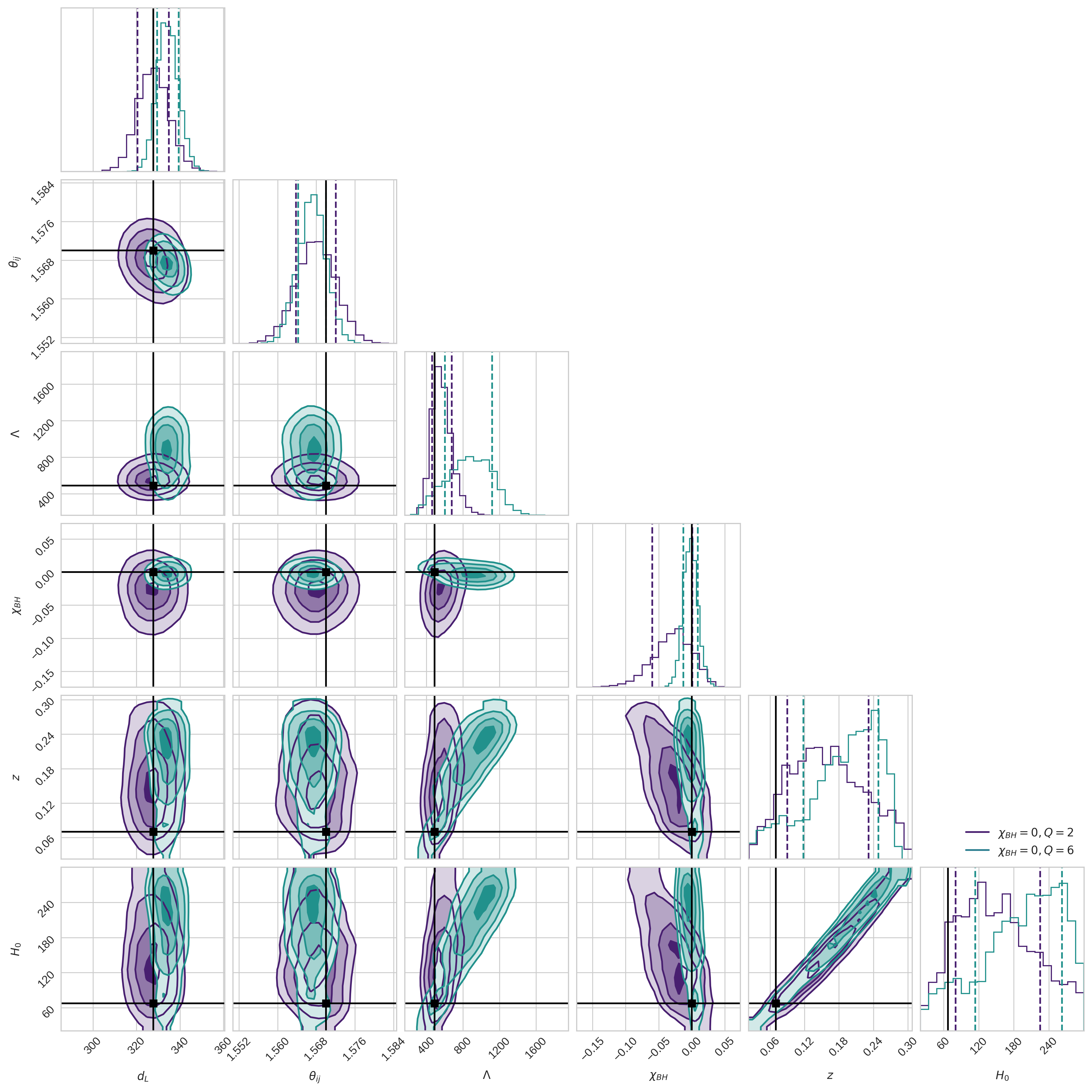}
\centering
\caption{\textbf{(Single runs)}: The posterior probability density and the 68\% credible intervals of $\{d_L, \theta_{ij}, \Lambda, \chi_{BH}, z, H_0\}$ as measured for $Q=2$ (purple) and $Q=6$ (green) systems located at $z=0.07$. The plot shows the results of direct sampling with pycbc.
}
\label{fig:fixedEOS}
\end{figure*}

\begin{figure*}
\includegraphics[width=0.9
\linewidth]{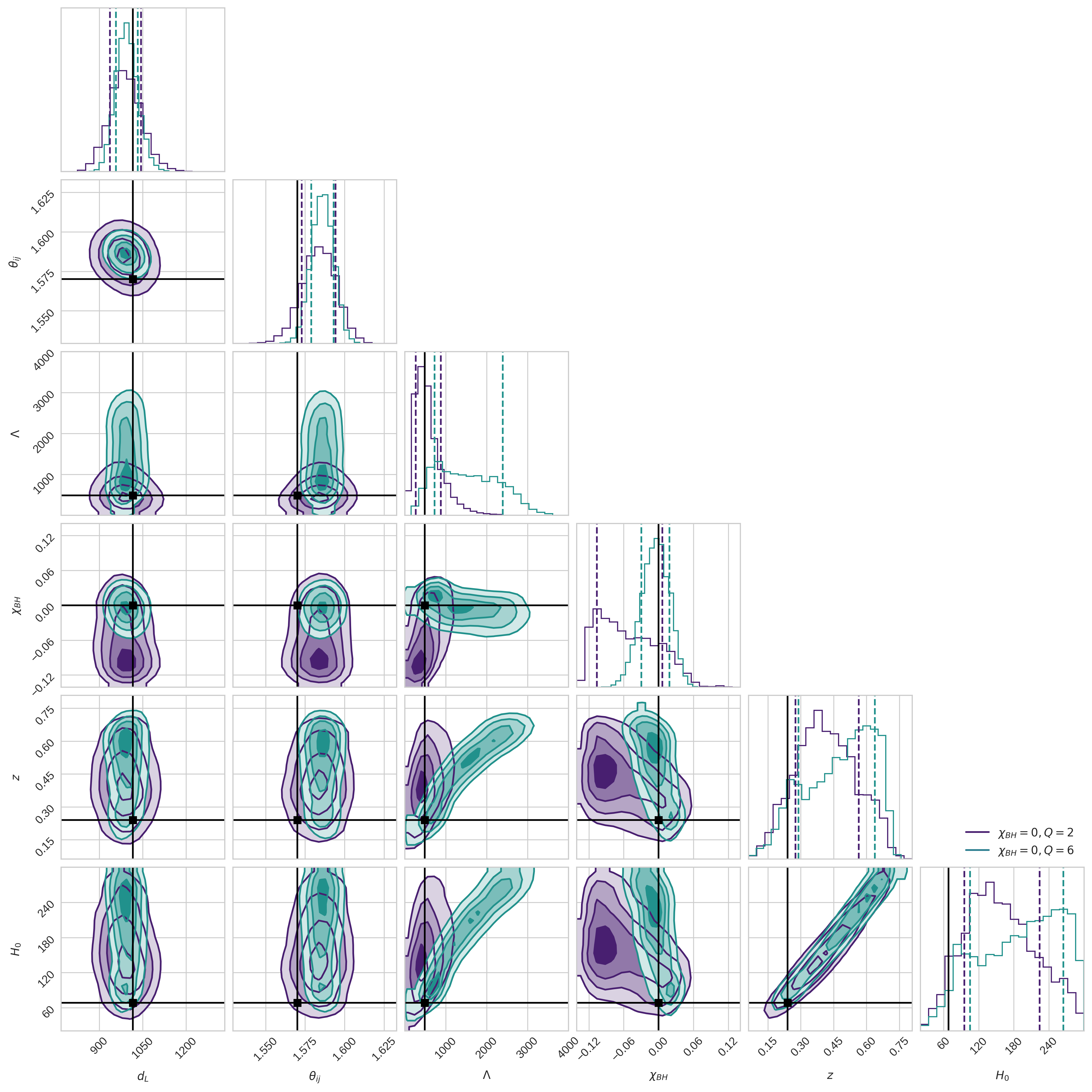}
\centering
\caption{\textbf{(Single runs)}: The posterior probability density and the 68\% credible intervals of $\{d_L, \theta_{ij}, \Lambda, \chi_{BH}, z, H_0\}$ as measured for $Q=2$ (purple) and $Q=6$ (green) systems located at $z=0.2$. The plot shows the results of direct sampling with pycbc.}
\label{fig:fixedEOS}
\end{figure*}
In order to see the degeneracy between sampled GW parameters and their effect on the overall inference, we have shown the full inference results for two of the considered examples, including the bounds on $H_0$ from these single events.For this section we are showing the results with the method of direct sampling of $H_0$, instead of the post-processing approach. This is necessary if we want to capture the degeneracy between the $\Lambda,z,H_0$ parameters all in one plot. We confirm that both of the methods result in similar constraints on the parameters. 

Our results for the inference of the redshift parameter are comparable with the similar bounds found by studying single BNS systems at $z<1$~\citep{PhysRevLett.108.091101}. However, while the authors of ~\citet{PhysRevLett.108.091101} considered a $Q=1, m_{NS,s}=1.4 M_\odot$ BNS case, such a system does not represent a physical NSBH binary and hence we can not perform a one-to-one comparison with the results of~\citet{PhysRevLett.108.091101}.

\begin{figure}
\includegraphics[width=1\linewidth]{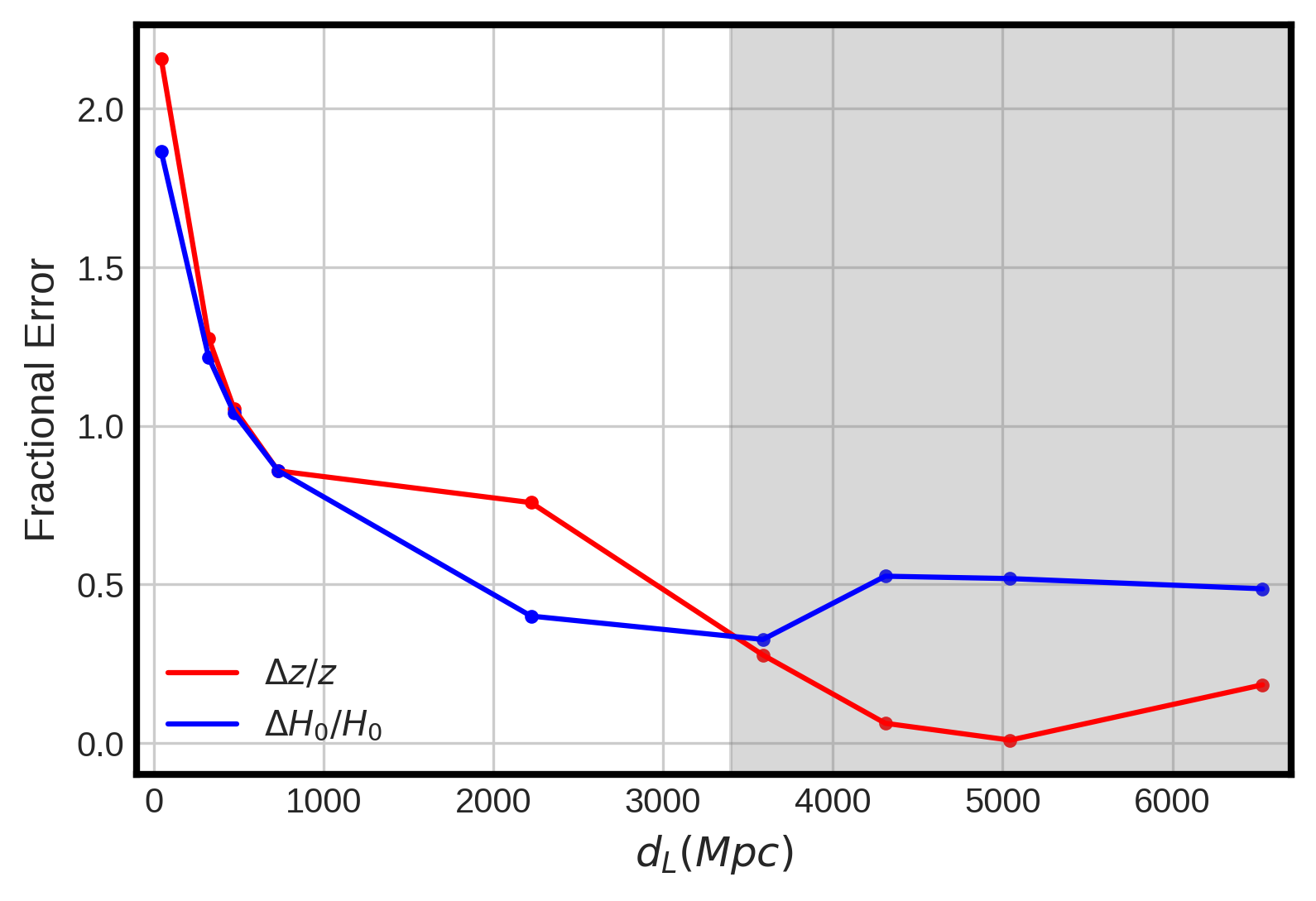}
\centering
\caption{The fractional statistical error in measurement of $z$ (red) and $H_0$ (blue) as a function of $d_L$ for a few representative parameter-estimation runs with $Q=2$. The injected parameters for these sample runs are same as those given in Sec.~\ref{sec:simulation}. The gray shaded region shows the threshold of SNR=30.}

\label{FracErr}
\end{figure}

We note that, even for highly-disruptive systems, the measurement of $H_{0}$ by individual events is affected by covariances between the parameters that result in a low precision, as well as hidden systematic uncertainties. This can be seen, for instance, for the low-redshift points of Fig.~\ref{FracErr}, where the error on $H_0$ inference is following the trend of $z$ error, rather than being small.
Also, as shown in Fig.~\ref{FracErr}, we clearly see that the ability to constrain $H_0$ from single events is limited at high distances and saturates at a certain error limit. This follows from the limitation on the inference of $z$, which is mainly tied to the uncertainties in inferring $\Lambda$,as well as $d_L$.
A similar analysis in~\cite{Chatterjee:2021xrm} as well as in~\cite{2022arXiv220311756G} shows the same limitations on the $H_0$ inference at high distances, for the case of BNS systems, even though the modelling approaches are different from what we consider. Although not shown here, we have also found that fixing the choice of the EoS parameter does not significantly improve the fractional error estimates of $H_0$ and redshift at high distances.

Overall, in the case of multiple events, the statistical uncertainty is expected to decrease as inverse square root of number of events yet this does not affect the systematic uncertainties and general limitations. We note again that, we have assumed that calibration errors will be negligible in the next generation GW detectors, yet this assumption should be revised in future works.
\subsection{Stacked events: $H_0$ detectability}

The next generation detectors, such as CE and ET, are expected to detect tens of thousands of NSBH events per year and hence can provide precise statistical measurement of $H_0$. Here we stack the individual $H_0$ measurements of simulated synthetic NSBH mergers as described in Sec.~\ref{sec:simulation}.
The combined $H_0$ PDF is found by multiplying the single PDFs of eq.~\ref{H0PDF} such that:
\begin{equation}\label{stack}
    P(H_0| \{\mathbf{x}_1,...\mathbf{x}_{N}\}) = \prod_{i=1}^{N} P(H_0|\mathbf{x}_{i})=P_0 (H_0) \mathcal{L}_{i}(H_0)\,,
\end{equation}
where we use an overall flat prior $P(H_0)$ on the stacked $H_0$ to get the stacked PDF.

\begin{figure}
\includegraphics[width=1.01\linewidth]{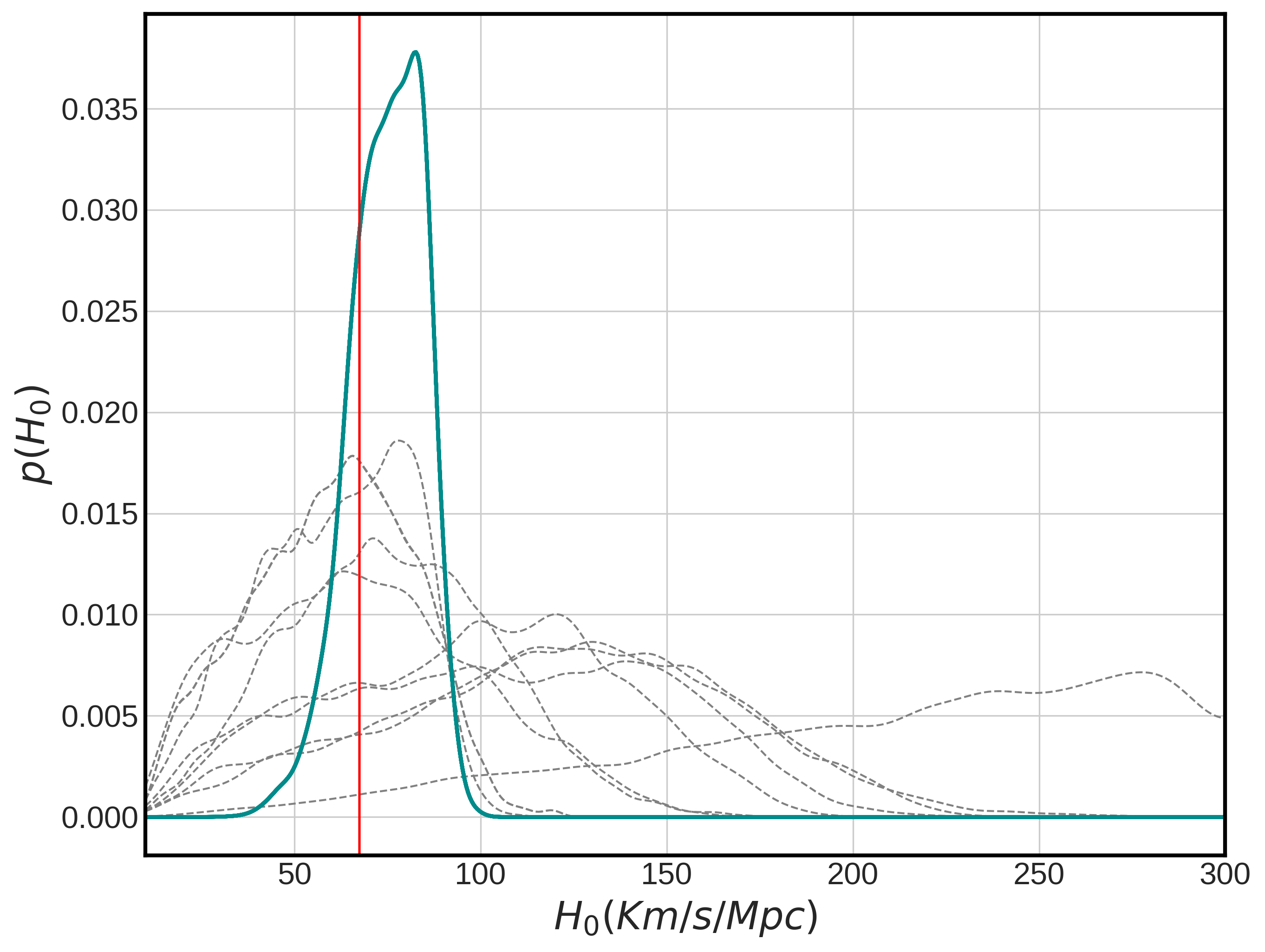}
\includegraphics[width=1.01\linewidth]{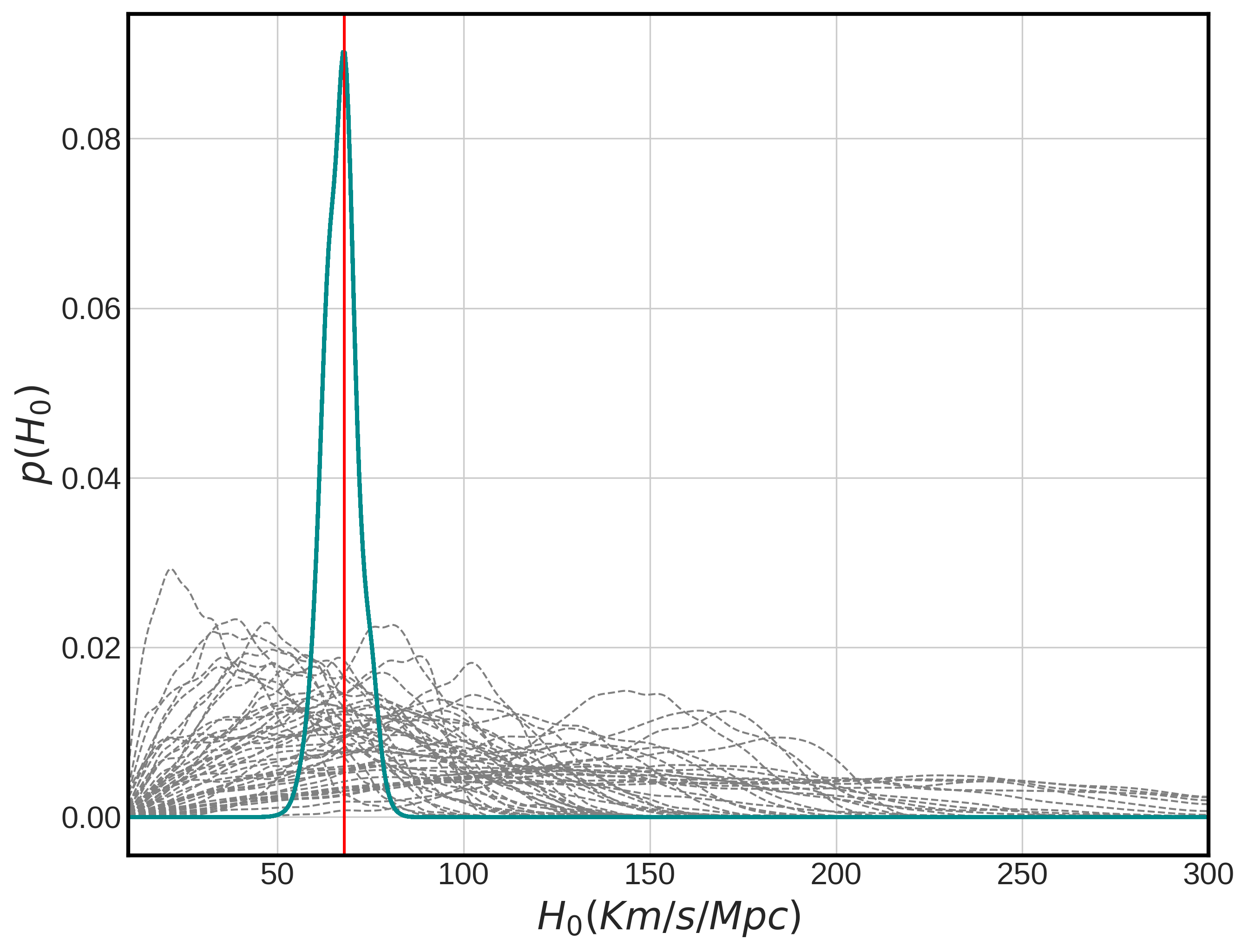}
\includegraphics[width=1.01\linewidth]{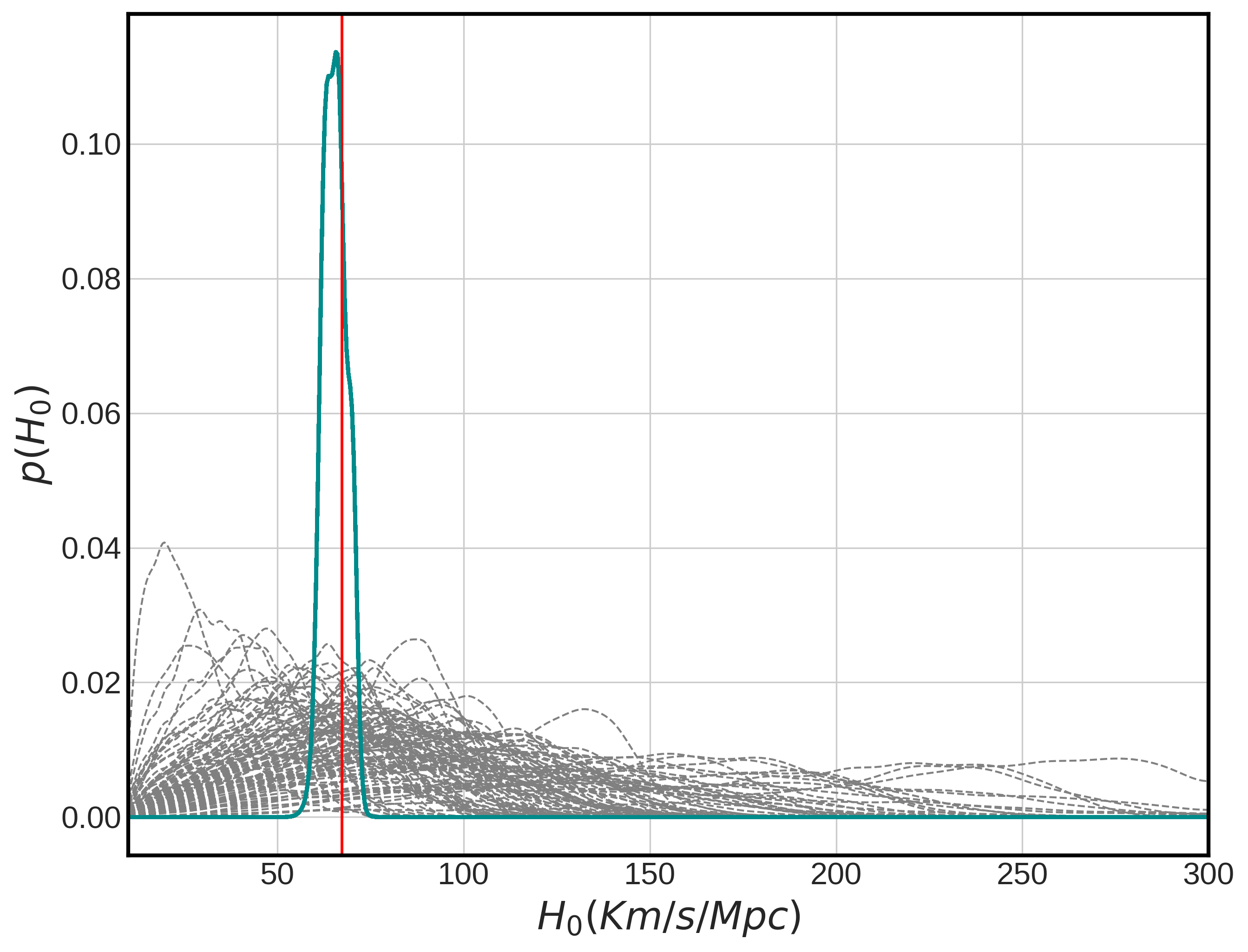}
\centering
\caption{The single and stacked H0 PDFs for simulated NSBH mergers in the ET+CEs detector era with $N=10,\,50,\,100$ simulated events. The vertical line corresponds to the true value of $H_0=67.4\, km/s/Mpc$. The estimated precision on $H_0$ are $\approx 13\%,\,6.6\%$ and $4\%$ for the $N=10,\,50,\,100$ case, respectively.}
\label{fig:LambdaStack}
\end{figure}

The results are shown in Fig.~\ref{fig:LambdaStack} for the different catalogues considered. The gray dashed lines show the individual PDFs and the green line shows the stacked PDF which is peaked at the injected $H_0$ value. We find that having $N=10, 50$ and  $100$, the $H_0$ can be measured with precision of $\approx 13\%,\, 6.6\%$ and $4\%$ (at $68\%$ credible interval). This precision is in agreement with the $\sqrt{N}$ scaling of the relative errors 
, which can be used as a first order estimate for uncertainties in a catalogue of events (\textit{cf.}~\citet{Mortlock:2018azx}).\\
We note that, in all the catalogues considered, the non-disruptive NSBH systems are the most dominant as they naturally cover a broader range of the parameter space once we generate the population parameters uniformly. This may not necessarily be the case for more realistic population models of NSBH systems (\textit{cf.}~\cite{Broekgaarden:2021iew,2019MNRAS.487....2M}). If a population allows for more disruptive mergers (or that we end up detecting more of these systems) \citep{Mapelli:2018wys}, we would naturally expect more stringent constraints on $\Lambda$-related parameters.  Also, note that we are not incorporating any GW detector selection effects in this analysis (\textit{cf.}~\citet{2019ApJ...882L..24A,2020PhRvD.102j3020G}, and references therein, for possible methods on including the selection biases.)

Our results for $N=100$ case show the same order of inference precision to that of a similar study performed on BNS mergers in the CE detector era~\citep{Chatterjee:2021xrm}. There the authors report a $2\%$ precision
in the measurement of $H_0$ with $N=\mathcal{O}(10^3)$ BNS events as seen by one CE detector, using the universal binary-Love relations to model $\Lambda$.


\section{Conclusions}\label{sec:conclustions}
We have analysed the prospects of inferring $H_0$ from the GW data of NSBH events without considering EM counterparts, based on the foundational idea that NS tidal effects can be exploited to break the redshift degeneracy of the GW waveforms, as proposed by~\cite{PhysRevLett.108.091101}. 
In this paper, we model the NS tidal parameter based on the latest multi-messenger constraints on the NS matter effects from NICER + LIGO/Virgo/KAGRA, hence modelling the $\Lambda$ parameter solely based on empirically viable EoS choices.
Our analysis is done on both single NSBH systems, as well as their synthetic catalogues, in the ET+CE detector era.

The unique merger phenomenology of NSBH systems allows for highly disruptive events that are key candidates for the precise measurement of NS tidal parameter, once they merge within the sensitivity band of the GW detectors considered. In the case of single events, by comparing some examples of disruptive and non-disruptive mergers, we show that in-band mergers of disruptive systems can constrain $\Lambda$ with $\approx$ $36\%$ precision, while this number increases to $\approx 50\%$ for the in-band mergers of non-disruptive systems. In the case of a system merging outside the detector's sensitivity bound, the disruptive mergers result in a mostly uninformative precision of $\approx 95\%$ and the non-disruptive mergers lead to a totally uninformative constraints on $\Lambda$.
Note however, that the specific values of these ranges are highly dependent on the specific initial signal configurations and may change by comparing signals with different e.g., duration, distance, inclination and sky location.

The overall improvement in $H_0$ inference is, however, limited due to the fundamental degeneracy between the redshift and $H_{0}$ parameter. Importantly, we find that the precision at which the redshift can be inferred with this approach, is not strong enough to result in unbiased and highly constrained measurements of $H_0$ from single events. This retains the need for accurate localization of systems in order to have a highly accurate redshift (and $H_0$) estimation. Also, we realise that the measurement of the redshift (and hence the $H_{0}$) with this approach is strongly affected by the details of spin precession and tidal distortion modelling in GW waveforms, highlighting the need for improved GW template models of NSBH mergers. 

In order to improve against this fundamental limitation of $H_0$ inference, we analyse synthetic sets of NSBH GW signals with samples of $N= 10,50,100$ events, with the population being generated by uniform sampling of all parameters. Our results show that for the ET + CS detector era, this method can result in unbiased $13-4\%$ precision in the measurement of $H_0$, once the same waveform model is used for the injection and recovery of the signals.

Future detailed analysis following this work can be done by including the effect of orbital precession in the analyzed signals. This is important as systems with large retrograde BH spins can result in significant orbital precession. We also note that, due to the intrinsic limitation of the waveform model Phenom-NSBH to $m_{NS,d}\ll 3M_\odot$, the detectable NS population depends on the redshift of the event. This modeling results in NSs with high masses to be less dominant at high redshifts and also that the lightest NSs (i.e., $ m_{NS,s} \approx 1 M_\odot$) can not be located at redshifts higher than $z=2$, which is a serious limiting factor in the analysis of these events in the next generation detector era. Also, including higher than quadrupole multipole moments in the waveform modelling is shown to be necessary for precise
parameter measurements for systems with $Q>3$~\citet{2020PhRvD.101j3004K}.
Another important avenue of improvement is to study more realistic NSBH population models. In principle, population models predicting more number of disruptive mergers would result in more stringent constraints on $H_0$.

\section{Acknowledgement}
We thank Stephen Feeney, Daniel Mortlock, Uddipta Bhardwaj and Suvodip Mukherjee for useful discussions and suggestions, and Archisman Gosh for comments on the manuscript. B.S., G.R. and S.M.N. are grateful for financial support from the Nederlandse Organisatie voor Wetenschappelijk Onderzoek (NWO) through the Projectruimte and VIDI grants (Nissanke). T.H. and S.M.N. also acknowledges financial support from the NWO sector plan. F.F. gratefully acknowledges support from the DOE through grant DE-SC0020435, and from NASA through grants 80NSSC18K0565 and 80NSSC22K0719.

\label{lastpage}

\begin{thebibliography}{}
\makeatletter
\relax
\def\mn@urlcharsother{\let\do\@makeother \do\$\do\&\do\#\do\^\do\_\do\%\do\~}
\def\mn@doi{\begingroup\mn@urlcharsother \@ifnextchar [ {\mn@doi@}
  {\mn@doi@[]}}
\def\mn@doi@[#1]#2{\def\@tempa{#1}\ifx\@tempa\@empty \href
  {http://dx.doi.org/#2} {doi:#2}\else \href {http://dx.doi.org/#2} {#1}\fi
  \endgroup}
\def\mn@eprint#1#2{\mn@eprint@#1:#2::\@nil}
\def\mn@eprint@arXiv#1{\href {http://arxiv.org/abs/#1} {{\tt arXiv:#1}}}
\def\mn@eprint@dblp#1{\href {http://dblp.uni-trier.de/rec/bibtex/#1.xml}
  {dblp:#1}}
\def\mn@eprint@#1:#2:#3:#4\@nil{\def\@tempa {#1}\def\@tempb {#2}\def\@tempc
  {#3}\ifx \@tempc \@empty \let \@tempc \@tempb \let \@tempb \@tempa \fi \ifx
  \@tempb \@empty \def\@tempb {arXiv}\fi \@ifundefined
  {mn@eprint@\@tempb}{\@tempb:\@tempc}{\expandafter \expandafter \csname
  mn@eprint@\@tempb\endcsname \expandafter{\@tempc}}}

\bibitem[\protect\citeauthoryear{Abbott et~al.}{Abbott
  et~al.}{2017a}]{Abbott:2017xzu}
Abbott B.~P.,  et~al., 2017a, \mn@doi [Nature] {10.1038/nature24471}, 551, 85

\bibitem[\protect\citeauthoryear{Abbott et~al.}{Abbott
  et~al.}{2017b}]{LIGOScientific:2017zic}
Abbott B.~P.,  et~al., 2017b, \mn@doi [Astrophys. J. Lett.]
  {10.3847/2041-8213/aa920c}, 848, L13

\bibitem[\protect\citeauthoryear{Abbott et~al.,}{Abbott
  et~al.}{2019a}]{PhysRevX.9.031040}
Abbott B.~P.,  et~al., 2019a, \mn@doi [Phys. Rev. X]
  {10.1103/PhysRevX.9.031040}, 9, 031040

\bibitem[\protect\citeauthoryear{{Abbott} et~al.,}{{Abbott}
  et~al.}{2019b}]{2019ApJ...882L..24A}
{Abbott} B.~P.,  et~al., 2019b, \mn@doi [\apjl] {10.3847/2041-8213/ab3800},
  \href {https://ui.adsabs.harvard.edu/abs/2019ApJ...882L..24A} {882, L24}

\bibitem[\protect\citeauthoryear{Abbott et~al.}{Abbott
  et~al.}{2021}]{LIGOScientific:2021qlt}
Abbott R.,  et~al., 2021, \mn@doi [Astrophys. J. Lett.]
  {10.3847/2041-8213/ac082e}, 915, L5

\bibitem[\protect\citeauthoryear{Aghanim et~al.}{Aghanim
  et~al.}{2020}]{Planck:2018vyg}
Aghanim N.,  et~al., 2020, \mn@doi [Astron. Astrophys.]
  {10.1051/0004-6361/201833910}, 641, A6

\bibitem[\protect\citeauthoryear{{Al-Mamun} et~al.,}{{Al-Mamun}
  et~al.}{2021}]{Almamun21}
{Al-Mamun} M.,  et~al., 2021, \mn@doi [\prl] {10.1103/PhysRevLett.126.061101},
  \href {https://ui.adsabs.harvard.edu/abs/2021PhRvL.126f1101A} {126, 061101}

\bibitem[\protect\citeauthoryear{Biwer, Capano, De, Cabero, Brown, Nitz  \&
  Raymond}{Biwer et~al.}{2019}]{Biwer:2018osg}
Biwer C.~M.,  Capano C.~D.,  De S.,  Cabero M.,  Brown D.~A.,  Nitz A.~H.,
  Raymond V.,  2019, \mn@doi [Publ. Astron. Soc. Pac.]
  {10.1088/1538-3873/aaef0b}, 131, 024503

\bibitem[\protect\citeauthoryear{{Boersma} \& {van Leeuwen}}{{Boersma} \& {van
  Leeuwen}}{2022}]{Boersma:2022yoz}
{Boersma} O.~M.,  {van Leeuwen} J.,  2022, arXiv e-prints, \href
  {https://ui.adsabs.harvard.edu/abs/2022arXiv220202181B} {p. arXiv:2202.02181}

\bibitem[\protect\citeauthoryear{{Borhanian} \& {Sathyaprakash}}{{Borhanian} \&
  {Sathyaprakash}}{2022}]{2022arXiv220211048B}
{Borhanian} S.,  {Sathyaprakash} B.~S.,  2022, arXiv e-prints, \href
  {https://ui.adsabs.harvard.edu/abs/2022arXiv220211048B} {p. arXiv:2202.11048}

\bibitem[\protect\citeauthoryear{{Borhanian}, {Dhani}, {Gupta}, {Arun}  \&
  {Sathyaprakash}}{{Borhanian} et~al.}{2020}]{2020arXiv200702883B}
{Borhanian} S.,  {Dhani} A.,  {Gupta} A.,  {Arun} K.~G.,   {Sathyaprakash}
  B.~S.,  2020, arXiv e-prints, \href
  {https://ui.adsabs.harvard.edu/abs/2020arXiv200702883B} {p. arXiv:2007.02883}

\bibitem[\protect\citeauthoryear{{Broekgaarden} et~al.,}{{Broekgaarden}
  et~al.}{2021a}]{Broekgaarden21}
{Broekgaarden} F.~S.,  et~al., 2021a, \mn@doi [\mnras]
  {10.1093/mnras/stab2716}, \href
  {https://ui.adsabs.harvard.edu/abs/2021MNRAS.508.5028B} {508, 5028}

\bibitem[\protect\citeauthoryear{Broekgaarden et~al.,}{Broekgaarden
  et~al.}{2021b}]{Broekgaarden:2021iew}
Broekgaarden F.~S.,  et~al., 2021b, \mn@doi [Mon. Not. Roy. Astron. Soc.]
  {10.1093/mnras/stab2716}, 508, 5028

\bibitem[\protect\citeauthoryear{{Buchner} et~al.,}{{Buchner}
  et~al.}{2014}]{2014A&A...564A.125B}
{Buchner} J.,  et~al., 2014, \mn@doi [\aap] {10.1051/0004-6361/201322971},
  \href {https://ui.adsabs.harvard.edu/abs/2014A&A...564A.125B} {564, A125}

\bibitem[\protect\citeauthoryear{{Capano} et~al.,}{{Capano}
  et~al.}{2020}]{Capano:2019eae}
{Capano} C.~D.,  et~al., 2020, \mn@doi [Nature Astronomy]
  {10.1038/s41550-020-1014-6}, \href
  {https://ui.adsabs.harvard.edu/abs/2020NatAs...4..625C} {4, 625}

\bibitem[\protect\citeauthoryear{{Chase} et~al.,}{{Chase}
  et~al.}{2022}]{Chase22}
{Chase} E.~A.,  et~al., 2022, \mn@doi [\apj] {10.3847/1538-4357/ac3d25}, \href
  {https://ui.adsabs.harvard.edu/abs/2022ApJ...927..163C} {927, 163}

\bibitem[\protect\citeauthoryear{{Chatterjee}, {Hegade K.~R.}, {Holder},
  {Holz}, {Perkins}, {Yagi}  \& {Yunes}}{{Chatterjee}
  et~al.}{2021}]{Chatterjee:2021xrm}
{Chatterjee} D.,  {Hegade K.~R.} A.,  {Holder} G.,  {Holz} D.~E.,  {Perkins}
  S.,  {Yagi} K.,   {Yunes} N.,  2021, \mn@doi [\prd]
  {10.1103/PhysRevD.104.083528}, \href
  {https://ui.adsabs.harvard.edu/abs/2021PhRvD.104h3528C} {104, 083528}

\bibitem[\protect\citeauthoryear{Chen, Fishbach  \& Holz}{Chen
  et~al.}{2018}]{Chen:2017rfc}
Chen H.-Y.,  Fishbach M.,   Holz D.~E.,  2018, \mn@doi [Nature]
  {10.1038/s41586-018-0606-0}, 562, 545

\bibitem[\protect\citeauthoryear{{Chen}, {Haster}, {Vitale}, {Farr}  \&
  {Isi}}{{Chen} et~al.}{2020}]{Chen:2020gek}
{Chen} H.-Y.,  {Haster} C.-J.,  {Vitale} S.,  {Farr} W.~M.,   {Isi} M.,  2020,
  arXiv e-prints, \href {https://ui.adsabs.harvard.edu/abs/2020arXiv200914057C}
  {p. arXiv:2009.14057}

\bibitem[\protect\citeauthoryear{{Cigarr{\'a}n D{\'\i}az} \&
  {Mukherjee}}{{Cigarr{\'a}n D{\'\i}az} \&
  {Mukherjee}}{2022}]{2022MNRAS.511.2782C}
{Cigarr{\'a}n D{\'\i}az} C.,  {Mukherjee} S.,  2022, \mn@doi [\mnras]
  {10.1093/mnras/stac208}, \href
  {https://ui.adsabs.harvard.edu/abs/2022MNRAS.511.2782C} {511, 2782}

\bibitem[\protect\citeauthoryear{{Cutler} \& {Flanagan}}{{Cutler} \&
  {Flanagan}}{1994}]{1994PhRvD..49.2658C}
{Cutler} C.,  {Flanagan} {\'E}.~E.,  1994, \mn@doi [\prd]
  {10.1103/PhysRevD.49.2658}, \href
  {https://ui.adsabs.harvard.edu/abs/1994PhRvD..49.2658C} {49, 2658}

\bibitem[\protect\citeauthoryear{Dalal, Holz, Hughes  \& Jain}{Dalal
  et~al.}{2006}]{PhysRevD.74.063006}
Dalal N.,  Holz D.~E.,  Hughes S.~A.,   Jain B.,  2006, \mn@doi [Phys. Rev. D]
  {10.1103/PhysRevD.74.063006}, 74, 063006

\bibitem[\protect\citeauthoryear{Deaton et~al.,}{Deaton
  et~al.}{2013}]{Deaton:2013sla}
Deaton M.~B.,  et~al., 2013, \mn@doi [Astrophys. J.]
  {10.1088/0004-637X/776/1/47}, 776, 47

\bibitem[\protect\citeauthoryear{Del~Pozzo, Li  \& Messenger}{Del~Pozzo
  et~al.}{2017}]{PhysRevD.95.043502}
Del~Pozzo W.,  Li T. G.~F.,   Messenger C.,  2017, \mn@doi [Phys. Rev. D]
  {10.1103/PhysRevD.95.043502}, 95, 043502

\bibitem[\protect\citeauthoryear{Dietrich, Coughlin, Pang, Bulla, Heinzel,
  Issa, Tews  \& Antier}{Dietrich et~al.}{2020}]{Dietrich:2020efo}
Dietrich T.,  Coughlin M.~W.,  Pang P. T.~H.,  Bulla M.,  Heinzel J.,  Issa L.,
   Tews I.,   Antier S.,  2020, \mn@doi [Science] {10.1126/science.abb4317},
  370, 1450

\bibitem[\protect\citeauthoryear{Doneva \& Pappas}{Doneva \&
  Pappas}{2018}]{Doneva:2017jop}
Doneva D.~D.,  Pappas G.,  2018, \mn@doi [Astrophys. Space Sci. Libr.]
  {10.1007/978-3-319-97616-7_13}, 457, 737

\bibitem[\protect\citeauthoryear{{Essick}, {Landry}  \& {Holz}}{{Essick}
  et~al.}{2020}]{Essick:2019ldf}
{Essick} R.,  {Landry} P.,   {Holz} D.~E.,  2020, \mn@doi [\prd]
  {10.1103/PhysRevD.101.063007}, \href
  {https://ui.adsabs.harvard.edu/abs/2020PhRvD.101f3007E} {101, 063007}

\bibitem[\protect\citeauthoryear{Etienne, Liu, Shapiro  \& Baumgarte}{Etienne
  et~al.}{2009}]{Etienne:2008re}
Etienne Z.~B.,  Liu Y.~T.,  Shapiro S.~L.,   Baumgarte T.~W.,  2009, \mn@doi
  [Phys. Rev. D] {10.1103/PhysRevD.79.044024}, 79, 044024

\bibitem[\protect\citeauthoryear{{Evans} et~al.,}{{Evans}
  et~al.}{2021}]{Evans:2021gyd}
{Evans} M.,  et~al., 2021, arXiv e-prints, \href
  {https://ui.adsabs.harvard.edu/abs/2021arXiv210909882E} {p. arXiv:2109.09882}

\bibitem[\protect\citeauthoryear{{Farr}, {Fishbach}, {Ye}  \& {Holz}}{{Farr}
  et~al.}{2019}]{2019ApJ...883L..42F}
{Farr} W.~M.,  {Fishbach} M.,  {Ye} J.,   {Holz} D.~E.,  2019, \mn@doi [\apjl]
  {10.3847/2041-8213/ab4284}, \href
  {https://ui.adsabs.harvard.edu/abs/2019ApJ...883L..42F} {883, L42}

\bibitem[\protect\citeauthoryear{Feeney, Peiris, Williamson, Nissanke,
  Mortlock, Alsing  \& Scolnic}{Feeney et~al.}{2019}]{PhysRevLett.122.061105}
Feeney S.~M.,  Peiris H.~V.,  Williamson A.~R.,  Nissanke S.~M.,  Mortlock
  D.~J.,  Alsing J.,   Scolnic D.,  2019, \mn@doi [Phys. Rev. Lett.]
  {10.1103/PhysRevLett.122.061105}, 122, 061105

\bibitem[\protect\citeauthoryear{Feeney, Peiris, Nissanke  \& Mortlock}{Feeney
  et~al.}{2021}]{Feeney:2020kxk}
Feeney S.~M.,  Peiris H.~V.,  Nissanke S.~M.,   Mortlock D.~J.,  2021, \mn@doi
  [Phys. Rev. Lett.] {10.1103/PhysRevLett.126.171102}, 126, 171102

\bibitem[\protect\citeauthoryear{Ferrari, Gualtieri  \& Pannarale}{Ferrari
  et~al.}{2009}]{Ferrari:2008nr}
Ferrari V.,  Gualtieri L.,   Pannarale F.,  2009, \mn@doi [Class. Quant. Grav.]
  {10.1088/0264-9381/26/12/125004}, 26, 125004

\bibitem[\protect\citeauthoryear{Ferrari, Gualtieri  \& Pannarale}{Ferrari
  et~al.}{2010}]{PhysRevD.81.064026}
Ferrari V.,  Gualtieri L.,   Pannarale F.,  2010, \mn@doi [Phys. Rev. D]
  {10.1103/PhysRevD.81.064026}, 81, 064026

\bibitem[\protect\citeauthoryear{Fishbach et~al.}{Fishbach
  et~al.}{2019}]{Fishbach:2018gjp}
Fishbach M.,  et~al., 2019, \mn@doi [Astrophys. J. Lett.]
  {10.3847/2041-8213/aaf96e}, 871, L13

\bibitem[\protect\citeauthoryear{Flanagan}{Flanagan}{1998}]{Flanagan:1997fn}
Flanagan E.~E.,  1998, \mn@doi [Phys. Rev. D] {10.1103/PhysRevD.58.124030}, 58,
  124030

\bibitem[\protect\citeauthoryear{Foucart}{Foucart}{2012}]{PhysRevD.86.124007}
Foucart F.,  2012, \mn@doi [Phys. Rev. D] {10.1103/PhysRevD.86.124007}, 86,
  124007

\bibitem[\protect\citeauthoryear{Foucart}{Foucart}{2020}]{Foucart:2020ats}
Foucart F.,  2020, \mn@doi [Front. Astron. Space Sci.]
  {10.3389/fspas.2020.00046}, 7, 46

\bibitem[\protect\citeauthoryear{Foucart et~al.,}{Foucart
  et~al.}{2013a}]{Foucart:2012vn}
Foucart F.,  et~al., 2013a, \mn@doi [Phys. Rev. D]
  {10.1103/PhysRevD.87.084006}, 87, 084006

\bibitem[\protect\citeauthoryear{Foucart et~al.,}{Foucart
  et~al.}{2013b}]{Foucart:2013psa}
Foucart F.,  et~al., 2013b, \mn@doi [Phys. Rev. D]
  {10.1103/PhysRevD.88.064017}, 88, 064017

\bibitem[\protect\citeauthoryear{{Gayathri} et~al.,}{{Gayathri}
  et~al.}{2020}]{Gayathri:2020fbl}
{Gayathri} V.,  et~al., 2020, arXiv e-prints, \href
  {https://ui.adsabs.harvard.edu/abs/2020arXiv200914247G} {p. arXiv:2009.14247}

\bibitem[\protect\citeauthoryear{{Gendreau} et~al.,}{{Gendreau}
  et~al.}{2016}]{2016SPIE.9905E..1HG}
{Gendreau} K.~C.,  et~al., 2016, in {den Herder} J.-W.~A.,  {Takahashi} T.,
  {Bautz} M.,  eds,  Society of Photo-Optical Instrumentation Engineers (SPIE)
  Conference Series Vol. 9905, Space Telescopes and Instrumentation 2016:
  Ultraviolet to Gamma Ray. p. 99051H, \mn@doi{10.1117/12.2231304}

\bibitem[\protect\citeauthoryear{{Gerosa}, {Pratten}  \& {Vecchio}}{{Gerosa}
  et~al.}{2020}]{2020PhRvD.102j3020G}
{Gerosa} D.,  {Pratten} G.,   {Vecchio} A.,  2020, \mn@doi [\prd]
  {10.1103/PhysRevD.102.103020}, \href
  {https://ui.adsabs.harvard.edu/abs/2020PhRvD.102j3020G} {102, 103020}

\bibitem[\protect\citeauthoryear{{Ghosh}, {Biswas}  \& {Bose}}{{Ghosh}
  et~al.}{2022}]{2022arXiv220311756G}
{Ghosh} T.,  {Biswas} B.,   {Bose} S.,  2022, arXiv e-prints, \href
  {https://ui.adsabs.harvard.edu/abs/2022arXiv220311756G} {p. arXiv:2203.11756}

\bibitem[\protect\citeauthoryear{Gossan, Hall  \& Nissanke}{Gossan
  et~al.}{2022}]{Gossan:2021eqe}
Gossan S.~E.,  Hall E.~D.,   Nissanke S.~M.,  2022, \mn@doi [Astrophys. J.]
  {10.3847/1538-4357/ac4164}, 926, 231

\bibitem[\protect\citeauthoryear{{Gray}, {Messenger}  \& {Veitch}}{{Gray}
  et~al.}{2022}]{2022MNRAS.512.1127G}
{Gray} R.,  {Messenger} C.,   {Veitch} J.,  2022, \mn@doi [\mnras]
  {10.1093/mnras/stac366}, \href
  {https://ui.adsabs.harvard.edu/abs/2022MNRAS.512.1127G} {512, 1127}

\bibitem[\protect\citeauthoryear{{Greif}, {Raaijmakers}, {Hebeler}, {Schwenk}
  \& {Watts}}{{Greif} et~al.}{2019}]{Greif19}
{Greif} S.~K.,  {Raaijmakers} G.,  {Hebeler} K.,  {Schwenk} A.,   {Watts}
  A.~L.,  2019, \mn@doi [MNRAS] {10.1093/mnras/stz654}, \href
  {http://adsabs.harvard.edu/abs/2019MNRAS.485.5363G} {485, 5363}

\bibitem[\protect\citeauthoryear{Guerra~Chaves \& Hinderer}{Guerra~Chaves \&
  Hinderer}{2019}]{GuerraChaves:2019foa}
Guerra~Chaves A.,  Hinderer T.,  2019, \mn@doi [J. Phys. G]
  {10.1088/1361-6471/ab45be}, 46, 123002

\bibitem[\protect\citeauthoryear{{Hebeler}, {Lattimer}, {Pethick}  \&
  {Schwenk}}{{Hebeler} et~al.}{2013}]{Hebeler13}
{Hebeler} K.,  {Lattimer} J.~M.,  {Pethick} C.~J.,   {Schwenk} A.,  2013,
  \mn@doi [\apj] {10.1088/0004-637X/773/1/11}, \href
  {https://ui.adsabs.harvard.edu/abs/2013ApJ...773...11H} {773, 11}

\bibitem[\protect\citeauthoryear{Hinderer, Lackey, Lang  \& Read}{Hinderer
  et~al.}{2010a}]{Hinderer:2009ca}
Hinderer T.,  Lackey B.~D.,  Lang R.~N.,   Read J.~S.,  2010a, \mn@doi [Phys.
  Rev. D] {10.1103/PhysRevD.81.123016}, 81, 123016

\bibitem[\protect\citeauthoryear{Hinderer, Lackey, Lang  \& Read}{Hinderer
  et~al.}{2010b}]{PhysRevD.81.123016}
Hinderer T.,  Lackey B.~D.,  Lang R.~N.,   Read J.~S.,  2010b, \mn@doi [Phys.
  Rev. D] {10.1103/PhysRevD.81.123016}, 81, 123016

\bibitem[\protect\citeauthoryear{Holz \& Hughes}{Holz \&
  Hughes}{2005}]{Holz:2005df}
Holz D.~E.,  Hughes S.~A.,  2005, \mn@doi [Astrophys. J.] {10.1086/431341},
  629, 15

\bibitem[\protect\citeauthoryear{Hotokezaka, Nakar, Gottlieb, Nissanke, Masuda,
  Hallinan, Mooley  \& Deller}{Hotokezaka et~al.}{2019}]{Hotokezaka:2018dfi}
Hotokezaka K.,  Nakar E.,  Gottlieb O.,  Nissanke S.,  Masuda K.,  Hallinan G.,
   Mooley K.~P.,   Deller A.~T.,  2019, \mn@doi [Nature Astron.]
  {10.1038/s41550-019-0820-1}, 3, 940

\bibitem[\protect\citeauthoryear{{Huang}, {Chen}, {Haster}, {Sun}, {Vitale}  \&
  {Kissel}}{{Huang} et~al.}{2022}]{2022arXiv220403614H}
{Huang} Y.,  {Chen} H.-Y.,  {Haster} C.-J.,  {Sun} L.,  {Vitale} S.,   {Kissel}
  J.,  2022, arXiv e-prints, \href
  {https://ui.adsabs.harvard.edu/abs/2022arXiv220403614H} {p. arXiv:2204.03614}

\bibitem[\protect\citeauthoryear{Hubble}{Hubble}{1929}]{Hubble168}
Hubble E.,  1929, \mn@doi [Proceedings of the National Academy of Sciences]
  {10.1073/pnas.15.3.168}, 15, 168

\bibitem[\protect\citeauthoryear{{Huth} et~al.,}{{Huth} et~al.}{2022}]{Huth22}
{Huth} S.,  et~al., 2022, \mn@doi [\nat] {10.1038/s41586-022-04750-w}, \href
  {https://ui.adsabs.harvard.edu/abs/2022Natur.606..276H} {606, 276}

\bibitem[\protect\citeauthoryear{{Kalaghatgi}, {Hannam}  \&
  {Raymond}}{{Kalaghatgi} et~al.}{2020}]{2020PhRvD.101j3004K}
{Kalaghatgi} C.,  {Hannam} M.,   {Raymond} V.,  2020, \mn@doi [\prd]
  {10.1103/PhysRevD.101.103004}, \href
  {https://ui.adsabs.harvard.edu/abs/2020PhRvD.101j3004K} {101, 103004}

\bibitem[\protect\citeauthoryear{Kelley}{Kelley}{2021}]{Kelley2021}
Kelley L.~Z.,  2021, \mn@doi [Journal of Open Source Software]
  {10.21105/joss.02784}, 6, 2784

\bibitem[\protect\citeauthoryear{{LIGO Scientific Collaboration}}{{LIGO
  Scientific Collaboration}}{2018}]{lalsuite}
{LIGO Scientific Collaboration} 2018, {LIGO} {A}lgorithm {L}ibrary -
  {LALS}uite, free software (GPL), \mn@doi{10.7935/GT1W-FZ16}

\bibitem[\protect\citeauthoryear{Lackey, Kyutoku, Shibata, Brady  \&
  Friedman}{Lackey et~al.}{2014}]{PhysRevD.89.043009}
Lackey B.~D.,  Kyutoku K.,  Shibata M.,  Brady P.~R.,   Friedman J.~L.,  2014,
  \mn@doi [Phys. Rev. D] {10.1103/PhysRevD.89.043009}, 89, 043009

\bibitem[\protect\citeauthoryear{{Landry} \& {Essick}}{{Landry} \&
  {Essick}}{2019}]{Landry19}
{Landry} P.,  {Essick} R.,  2019, \mn@doi [\prd] {10.1103/PhysRevD.99.084049},
  \href {https://ui.adsabs.harvard.edu/abs/2019PhRvD..99h4049L} {99, 084049}

\bibitem[\protect\citeauthoryear{{Landry}, {Essick}  \&
  {Chatziioannou}}{{Landry} et~al.}{2020}]{Landry20}
{Landry} P.,  {Essick} R.,   {Chatziioannou} K.,  2020, \mn@doi [\prd]
  {10.1103/PhysRevD.101.123007}, \href
  {https://ui.adsabs.harvard.edu/abs/2020PhRvD.101l3007L} {101, 123007}

\bibitem[\protect\citeauthoryear{Lattimer \& Schramm}{Lattimer \&
  Schramm}{1974}]{Lattimer:1974slx}
Lattimer J.~M.,  Schramm D.~N.,  1974, \mn@doi [Astrophys. J. Lett.]
  {10.1086/181612}, 192, L145

\bibitem[\protect\citeauthoryear{{Legred}, {Chatziioannou}, {Essick}, {Han}  \&
  {Landry}}{{Legred} et~al.}{2021}]{LegredJ0740}
{Legred} I.,  {Chatziioannou} K.,  {Essick} R.,  {Han} S.,   {Landry} P.,
  2021, \mn@doi [\prd] {10.1103/PhysRevD.104.063003}, \href
  {https://ui.adsabs.harvard.edu/abs/2021PhRvD.104f3003L} {104, 063003}

\bibitem[\protect\citeauthoryear{{Lindblom}}{{Lindblom}}{2018}]{Lindblom18}
{Lindblom} L.,  2018, \mn@doi [\prd] {10.1103/PhysRevD.97.123019}, \href
  {https://ui.adsabs.harvard.edu/abs/2018PhRvD..97l3019L} {97, 123019}

\bibitem[\protect\citeauthoryear{Maggiore et~al.}{Maggiore
  et~al.}{2020}]{Maggiore:2019uih}
Maggiore M.,  et~al., 2020, \mn@doi [JCAP] {10.1088/1475-7516/2020/03/050}, 03,
  050

\bibitem[\protect\citeauthoryear{Mapelli \& Giacobbo}{Mapelli \&
  Giacobbo}{2018}]{Mapelli:2018wys}
Mapelli M.,  Giacobbo N.,  2018, \mn@doi [Mon. Not. Roy. Astron. Soc.]
  {10.1093/mnras/sty1613}, 479, 4391

\bibitem[\protect\citeauthoryear{{Mapelli}, {Giacobbo}, {Santoliquido}  \&
  {Artale}}{{Mapelli} et~al.}{2019}]{2019MNRAS.487....2M}
{Mapelli} M.,  {Giacobbo} N.,  {Santoliquido} F.,   {Artale} M.~C.,  2019,
  \mn@doi [\mnras] {10.1093/mnras/stz1150}, \href
  {https://ui.adsabs.harvard.edu/abs/2019MNRAS.487....2M} {487, 2}

\bibitem[\protect\citeauthoryear{{Mastrogiovanni} et~al.,}{{Mastrogiovanni}
  et~al.}{2021}]{2021PhRvD.104f2009M}
{Mastrogiovanni} S.,  et~al., 2021, \mn@doi [\prd]
  {10.1103/PhysRevD.104.062009}, \href
  {https://ui.adsabs.harvard.edu/abs/2021PhRvD.104f2009M} {104, 062009}

\bibitem[\protect\citeauthoryear{Matas et~al.}{Matas
  et~al.}{2020}]{Matas:2020wab}
Matas A.,  et~al., 2020, \mn@doi [Phys. Rev. D] {10.1103/PhysRevD.102.043023},
  102, 043023

\bibitem[\protect\citeauthoryear{Messenger \& Read}{Messenger \&
  Read}{2012}]{PhysRevLett.108.091101}
Messenger C.,  Read J.,  2012, \mn@doi [Phys. Rev. Lett.]
  {10.1103/PhysRevLett.108.091101}, 108, 091101

\bibitem[\protect\citeauthoryear{{Metzger} \& {Berger}}{{Metzger} \&
  {Berger}}{2012}]{2012ApJ...746...48M}
{Metzger} B.~D.,  {Berger} E.,  2012, \mn@doi [\apj]
  {10.1088/0004-637X/746/1/48}, \href
  {https://ui.adsabs.harvard.edu/abs/2012ApJ...746...48M} {746, 48}

\bibitem[\protect\citeauthoryear{{Miller} et~al.,}{{Miller}
  et~al.}{2019}]{MillerJ0030}
{Miller} M.~C.,  et~al., 2019, \mn@doi [\apjl] {10.3847/2041-8213/ab50c5},
  \href {https://ui.adsabs.harvard.edu/abs/2019ApJ...887L..24M} {887, L24}

\bibitem[\protect\citeauthoryear{{Miller} et~al.,}{{Miller}
  et~al.}{2021}]{MillerJ0740}
{Miller} M.~C.,  et~al., 2021, \mn@doi [\apjl] {10.3847/2041-8213/ac089b},
  \href {https://ui.adsabs.harvard.edu/abs/2021ApJ...918L..28M} {918, L28}

\bibitem[\protect\citeauthoryear{Mortlock, Feeney, Peiris, Williamson  \&
  Nissanke}{Mortlock et~al.}{2019a}]{PhysRevD.100.103523}
Mortlock D.~J.,  Feeney S.~M.,  Peiris H.~V.,  Williamson A.~R.,   Nissanke
  S.~M.,  2019a, \mn@doi [Phys. Rev. D] {10.1103/PhysRevD.100.103523}, 100,
  103523

\bibitem[\protect\citeauthoryear{Mortlock, Feeney, Peiris, Williamson  \&
  Nissanke}{Mortlock et~al.}{2019b}]{Mortlock:2018azx}
Mortlock D.~J.,  Feeney S.~M.,  Peiris H.~V.,  Williamson A.~R.,   Nissanke
  S.~M.,  2019b, \mn@doi [Phys. Rev. D] {10.1103/PhysRevD.100.103523}, 100,
  103523

\bibitem[\protect\citeauthoryear{{Mukherjee} et~al.,}{{Mukherjee}
  et~al.}{2020}]{Mukherjee:2020kki}
{Mukherjee} S.,  et~al., 2020, arXiv e-prints, \href
  {https://ui.adsabs.harvard.edu/abs/2020arXiv200914199M} {p. arXiv:2009.14199}

\bibitem[\protect\citeauthoryear{Mukherjee, Wandelt, Nissanke  \&
  Silvestri}{Mukherjee et~al.}{2021a}]{Mukherjee:2020hyn}
Mukherjee S.,  Wandelt B.~D.,  Nissanke S.~M.,   Silvestri A.,  2021a, \mn@doi
  [Phys. Rev. D] {10.1103/PhysRevD.103.043520}, 103, 043520

\bibitem[\protect\citeauthoryear{{Mukherjee}, {Lavaux}, {Bouchet}, {Jasche},
  {Wandelt}, {Nissanke}, {Leclercq}  \& {Hotokezaka}}{{Mukherjee}
  et~al.}{2021b}]{2021A&A...646A..65M}
{Mukherjee} S.,  {Lavaux} G.,  {Bouchet} F.~R.,  {Jasche} J.,  {Wandelt} B.~D.,
   {Nissanke} S.,  {Leclercq} F.,   {Hotokezaka} K.,  2021b, \mn@doi [\aap]
  {10.1051/0004-6361/201936724}, \href
  {https://ui.adsabs.harvard.edu/abs/2021A&A...646A..65M} {646, A65}

\bibitem[\protect\citeauthoryear{{Mukherjee}, {Krolewski}, {Wandelt}  \&
  {Silk}}{{Mukherjee} et~al.}{2022}]{2022arXiv220303643M}
{Mukherjee} S.,  {Krolewski} A.,  {Wandelt} B.~D.,   {Silk} J.,  2022, arXiv
  e-prints, \href {https://ui.adsabs.harvard.edu/abs/2022arXiv220303643M} {p.
  arXiv:2203.03643}

\bibitem[\protect\citeauthoryear{Nicolaou, Lahav, Lemos, Hartley  \&
  Braden}{Nicolaou et~al.}{2020}]{Nicolaou:2019cip}
Nicolaou C.,  Lahav O.,  Lemos P.,  Hartley W.,   Braden J.,  2020, \mn@doi
  [Mon. Not. Roy. Astron. Soc.] {10.1093/mnras/staa1120}, 495, 90

\bibitem[\protect\citeauthoryear{Nissanke, Holz, Hughes, Dalal  \&
  Sievers}{Nissanke et~al.}{2010}]{Nissanke:2009kt}
Nissanke S.,  Holz D.~E.,  Hughes S.~A.,  Dalal N.,   Sievers J.~L.,  2010,
  \mn@doi [Astrophys. J.] {10.1088/0004-637X/725/1/496}, 725, 496

\bibitem[\protect\citeauthoryear{{Nissanke}, {Holz}, {Dalal}, {Hughes},
  {Sievers}  \& {Hirata}}{{Nissanke} et~al.}{2013a}]{Nissanke:2013fka}
{Nissanke} S.,  {Holz} D.~E.,  {Dalal} N.,  {Hughes} S.~A.,  {Sievers} J.~L.,
  {Hirata} C.~M.,  2013a, arXiv e-prints, \href
  {https://ui.adsabs.harvard.edu/abs/2013arXiv1307.2638N} {p. arXiv:1307.2638}

\bibitem[\protect\citeauthoryear{{Nissanke}, {Kasliwal}  \&
  {Georgieva}}{{Nissanke} et~al.}{2013b}]{2013ApJ...767..124N}
{Nissanke} S.,  {Kasliwal} M.,   {Georgieva} A.,  2013b, \mn@doi [\apj]
  {10.1088/0004-637X/767/2/124}, \href
  {https://ui.adsabs.harvard.edu/abs/2013ApJ...767..124N} {767, 124}

\bibitem[\protect\citeauthoryear{{O'Boyle}, {Markakis}, {Stergioulas}  \&
  {Read}}{{O'Boyle} et~al.}{2020}]{Boyle20}
{O'Boyle} M.~F.,  {Markakis} C.,  {Stergioulas} N.,   {Read} J.~S.,  2020,
  \mn@doi [\prd] {10.1103/PhysRevD.102.083027}, \href
  {https://ui.adsabs.harvard.edu/abs/2020PhRvD.102h3027O} {102, 083027}

\bibitem[\protect\citeauthoryear{Oguri}{Oguri}{2016}]{PhysRevD.93.083511}
Oguri M.,  2016, \mn@doi [Phys. Rev. D] {10.1103/PhysRevD.93.083511}, 93,
  083511

\bibitem[\protect\citeauthoryear{Palmese et~al.}{Palmese
  et~al.}{2020}]{Palmese:2020aof}
Palmese A.,  et~al., 2020, \mn@doi [Astrophys. J. Lett.]
  {10.3847/2041-8213/abaeff}, 900, L33

\bibitem[\protect\citeauthoryear{{Pang}, {Tews}, {Coughlin}, {Bulla}, {Van Den
  Broeck}  \& {Dietrich}}{{Pang} et~al.}{2021b}]{Pang0740}
{Pang} P. T.~H.,  {Tews} I.,  {Coughlin} M.~W.,  {Bulla} M.,  {Van Den Broeck}
  C.,   {Dietrich} T.,  2021b, \mn@doi [\apj] {10.3847/1538-4357/ac19ab}, \href
  {https://ui.adsabs.harvard.edu/abs/2021ApJ...922...14P} {922, 14}

\bibitem[\protect\citeauthoryear{Pang, Tews, Coughlin, Bulla, Van Den~Broeck
  \& Dietrich}{Pang et~al.}{2021a}]{Pang:2021jta}
Pang P. T.~H.,  Tews I.,  Coughlin M.~W.,  Bulla M.,  Van Den~Broeck C.,
  Dietrich T.,  2021a, \mn@doi [Astrophys. J.] {10.3847/1538-4357/ac19ab}, 922,
  14

\bibitem[\protect\citeauthoryear{Pannarale, Rezzolla, Ohme  \& Read}{Pannarale
  et~al.}{2011}]{PhysRevD.84.104017}
Pannarale F.,  Rezzolla L.,  Ohme F.,   Read J.~S.,  2011, \mn@doi [Phys. Rev.
  D] {10.1103/PhysRevD.84.104017}, 84, 104017

\bibitem[\protect\citeauthoryear{Pannarale, Berti, Kyutoku, Lackey  \&
  Shibata}{Pannarale et~al.}{2015a}]{PhysRevD.92.081504}
Pannarale F.,  Berti E.,  Kyutoku K.,  Lackey B.~D.,   Shibata M.,  2015a,
  \mn@doi [Phys. Rev. D] {10.1103/PhysRevD.92.081504}, 92, 081504

\bibitem[\protect\citeauthoryear{Pannarale, Berti, Kyutoku, Lackey  \&
  Shibata}{Pannarale et~al.}{2015b}]{Pannarale:2015jka}
Pannarale F.,  Berti E.,  Kyutoku K.,  Lackey B.~D.,   Shibata M.,  2015b,
  \mn@doi [Phys. Rev. D] {10.1103/PhysRevD.92.084050}, 92, 084050

\bibitem[\protect\citeauthoryear{{Raaijmakers} et~al.,}{{Raaijmakers}
  et~al.}{2021a}]{Raaijmakers2021a}
{Raaijmakers} G.,  et~al., 2021a, \mn@doi [\apjl] {10.3847/2041-8213/ac089a},
  \href {https://ui.adsabs.harvard.edu/abs/2021ApJ...918L..29R} {918, L29}

\bibitem[\protect\citeauthoryear{Raaijmakers et~al.}{Raaijmakers
  et~al.}{2021b}]{Raaijmakers:2021slr}
Raaijmakers G.,  et~al., 2021b, \mn@doi [Astrophys. J.]
  {10.3847/1538-4357/ac222d}, 922, 269

\bibitem[\protect\citeauthoryear{Reitze et~al.}{Reitze
  et~al.}{2019}]{Reitze:2019iox}
Reitze D.,  et~al., 2019, Bull. Am. Astron. Soc., 51, 035

\bibitem[\protect\citeauthoryear{{Riess} et~al.,}{{Riess}
  et~al.}{2021}]{Riess:2021jrx}
{Riess} A.~G.,  et~al., 2021, arXiv e-prints, \href
  {https://ui.adsabs.harvard.edu/abs/2021arXiv211204510R} {p. arXiv:2112.04510}

\bibitem[\protect\citeauthoryear{{Riley} et~al.,}{{Riley}
  et~al.}{2019}]{RileyJ0030}
{Riley} T.~E.,  et~al., 2019, \mn@doi [\apjl] {10.3847/2041-8213/ab481c}, \href
  {https://ui.adsabs.harvard.edu/abs/2019ApJ...887L..21R} {887, L21}

\bibitem[\protect\citeauthoryear{{Riley} et~al.,}{{Riley}
  et~al.}{2021}]{RileyJ0740}
{Riley} T.~E.,  et~al., 2021, \mn@doi [\apjl] {10.3847/2041-8213/ac0a81}, \href
  {https://ui.adsabs.harvard.edu/abs/2021ApJ...918L..27R} {918, L27}

\bibitem[\protect\citeauthoryear{{Sathyaprakash} et~al.,}{{Sathyaprakash}
  et~al.}{2019}]{2019BAAS...51c.276S}
{Sathyaprakash} B.,  et~al., 2019, \baas, \href
  {https://ui.adsabs.harvard.edu/abs/2019BAAS...51c.276S} {51, 276}

\bibitem[\protect\citeauthoryear{{Schutz}}{{Schutz}}{1986}]{1986Natur.323..310S}
{Schutz} B.~F.,  1986, \mn@doi [\nat] {10.1038/323310a0}, \href
  {https://ui.adsabs.harvard.edu/abs/1986Natur.323..310S} {323, 310}

\bibitem[\protect\citeauthoryear{Seto \& Kyutoku}{Seto \&
  Kyutoku}{2018}]{Seto:2017swx}
Seto N.,  Kyutoku K.,  2018, \mn@doi [Mon. Not. Roy. Astron. Soc.]
  {10.1093/mnras/sty090}, 475, 4133

\bibitem[\protect\citeauthoryear{Shibata, Kyutoku, Yamamoto  \&
  Taniguchi}{Shibata et~al.}{2009}]{PhysRevD.79.044030}
Shibata M.,  Kyutoku K.,  Yamamoto T.,   Taniguchi K.,  2009, \mn@doi [Phys.
  Rev. D] {10.1103/PhysRevD.79.044030}, 79, 044030

\bibitem[\protect\citeauthoryear{Soares-Santos et~al.}{Soares-Santos
  et~al.}{2019a}]{Soares-Santos:2019irc}
Soares-Santos M.,  et~al., 2019a, \mn@doi [Astrophys. J. Lett.]
  {10.3847/2041-8213/ab14f1}, 876, L7

\bibitem[\protect\citeauthoryear{Soares-Santos et~al.}{Soares-Santos
  et~al.}{2019b}]{DES:2019ccw}
Soares-Santos M.,  et~al., 2019b, \mn@doi [Astrophys. J. Lett.]
  {10.3847/2041-8213/ab14f1}, 876, L7

\bibitem[\protect\citeauthoryear{Speagle}{Speagle}{2020}]{10.1093/mnras/staa278}
Speagle J.~S.,  2020, \mn@doi [Monthly Notices of the Royal Astronomical
  Society] {10.1093/mnras/staa278}, 493, 3132

\bibitem[\protect\citeauthoryear{Stephens, East  \& Pretorius}{Stephens
  et~al.}{2011}]{Stephens:2011as}
Stephens B.~C.,  East W.~E.,   Pretorius F.,  2011, \mn@doi [Astrophys. J.
  Lett.] {10.1088/2041-8205/737/1/L5}, 737, L5

\bibitem[\protect\citeauthoryear{{Takami}, {Rezzolla}  \& {Baiotti}}{{Takami}
  et~al.}{2014}]{2014PhRvL.113i1104T}
{Takami} K.,  {Rezzolla} L.,   {Baiotti} L.,  2014, \mn@doi [\prl]
  {10.1103/PhysRevLett.113.091104}, \href
  {https://ui.adsabs.harvard.edu/abs/2014PhRvL.113i1104T} {113, 091104}

\bibitem[\protect\citeauthoryear{Taylor, Gair  \& Mandel}{Taylor
  et~al.}{2012}]{PhysRevD.85.023535}
Taylor S.~R.,  Gair J.~R.,   Mandel I.,  2012, \mn@doi [Phys. Rev. D]
  {10.1103/PhysRevD.85.023535}, 85, 023535

\bibitem[\protect\citeauthoryear{{The LIGO Scientific Collaboration}, {the
  Virgo Collaboration}  \& {the KAGRA Collaboration}}{{The LIGO Scientific
  Collaboration} et~al.}{2021a}]{gwtc3_hubble}
{The LIGO Scientific Collaboration} {the Virgo Collaboration}  {the KAGRA
  Collaboration} 2021a, arXiv e-prints, \href
  {https://ui.adsabs.harvard.edu/abs/2021arXiv211103604T} {p. arXiv:2111.03604}

\bibitem[\protect\citeauthoryear{{The LIGO Scientific Collaboration}
  et~al.,}{{The LIGO Scientific Collaboration} et~al.}{2021b}]{gwtc3}
{The LIGO Scientific Collaboration} et~al., 2021b, arXiv e-prints, \href
  {https://ui.adsabs.harvard.edu/abs/2021arXiv211103606T} {p. arXiv:2111.03606}

\bibitem[\protect\citeauthoryear{Thompson, Fauchon-Jones, Khan, Nitoglia,
  Pannarale, Dietrich  \& Hannam}{Thompson et~al.}{2020}]{Thompson:2020nei}
Thompson J.~E.,  Fauchon-Jones E.,  Khan S.,  Nitoglia E.,  Pannarale F.,
  Dietrich T.,   Hannam M.,  2020, \mn@doi [Phys. Rev. D]
  {10.1103/PhysRevD.101.124059}, 101, 124059

\bibitem[\protect\citeauthoryear{Vallisneri}{Vallisneri}{2000}]{Vallisneri:1999nq}
Vallisneri M.,  2000, \mn@doi [Phys. Rev. Lett.] {10.1103/PhysRevLett.84.3519},
  84, 3519

\bibitem[\protect\citeauthoryear{Vasylyev \& Filippenko}{Vasylyev \&
  Filippenko}{2020}]{Vasylyev:2020hgb}
Vasylyev S.,  Filippenko A.,  2020, \mn@doi [Astrophys. J.]
  {10.3847/1538-4357/abb5f9}, 902, 149

\bibitem[\protect\citeauthoryear{Vines, Flanagan  \& Hinderer}{Vines
  et~al.}{2011}]{PhysRevD.83.084051}
Vines J.,  Flanagan E.~E.,   Hinderer T.,  2011, \mn@doi [Phys. Rev. D]
  {10.1103/PhysRevD.83.084051}, 83, 084051

\bibitem[\protect\citeauthoryear{Vitale \& Chen}{Vitale \&
  Chen}{2018a}]{PhysRevLett.121.021303}
Vitale S.,  Chen H.-Y.,  2018a, \mn@doi [Phys. Rev. Lett.]
  {10.1103/PhysRevLett.121.021303}, 121, 021303

\bibitem[\protect\citeauthoryear{Vitale \& Chen}{Vitale \&
  Chen}{2018b}]{Vitale:2018wlg}
Vitale S.,  Chen H.-Y.,  2018b, \mn@doi [Phys. Rev. Lett.]
  {10.1103/PhysRevLett.121.021303}, 121, 021303

\bibitem[\protect\citeauthoryear{Yagi \& Yunes}{Yagi \&
  Yunes}{2017}]{Yagi:2016bkt}
Yagi K.,  Yunes N.,  2017, \mn@doi [Phys. Rept.]
  {10.1016/j.physrep.2017.03.002}, 681, 1

\bibitem[\protect\citeauthoryear{{Zackay}, {Dai}  \& {Venumadhav}}{{Zackay}
  et~al.}{2018}]{Zackay:2018qdy}
{Zackay} B.,  {Dai} L.,   {Venumadhav} T.,  2018, arXiv e-prints, \href
  {https://ui.adsabs.harvard.edu/abs/2018arXiv180608792Z} {p. arXiv:1806.08792}

\makeatother
\end{thebibliography}
\end{document}